\documentclass[11pt]{article}

\usepackage[margin=1in]{geometry}
\usepackage[T1]{fontenc}
\usepackage[utf8]{inputenc}
\usepackage{lmodern}
\usepackage{pgf}
\usepackage{microtype}

\usepackage{caption}
\usepackage{subcaption}
\usepackage{comment}
\usepackage{booktabs}
\usepackage[authoryear]{natbib}
\usepackage{amsmath,amssymb,amsthm,mathtools}
\usepackage{threeparttable}
\usepackage{tabularx}
\usepackage{bbm}
\usepackage{array}
\usepackage{enumitem}
\usepackage{float}

\usepackage{graphicx}
\usepackage{xcolor}
\usepackage{hyperref}
\usepackage[nameinlink,noabbrev]{cleveref}

\usepackage{authblk}

\usepackage[ruled,vlined]{algorithm2e}
\SetKwInput{Input}{Inputs}
\SetKwInput{Output}{Output}
\newcommand{\AlgStep}[1]{\textbf{Step #1}}

\usepackage{listings}

\hypersetup{
  colorlinks=true,
  linkcolor=blue,
  citecolor=blue,
  urlcolor=blue
}

\newtheorem{theorem}{Theorem}

\newcommand{\RR}{\mathbb{R}}
\newcommand{\I}{\mathbbm{1}}
\newcommand{\Tstar}{T^{\ast}}

\newcommand{\Rset}{\mathcal{R}}
\newcommand{\Dland}{\mathcal{D}_{\ell}}

\newcommand{\figplaceholder}[4][]{%
\begin{figure}[t]
  \centering
  \IfFileExists{#2}{%
    \includegraphics[#1]{#2}%
  }{%
    \fbox{\parbox{0.9\linewidth}{\centering missing figure file: \texttt{#2}}}%
  }%
  \caption{#3}
  \label{#4}
\end{figure}
}

\newcommand{\covcell}[1]{%
\pgfmathtruncatemacro{\isgood}{abs(#1-0.90) <= 0.0201 ? 1 : 0}%
\ifnum\isgood=1 \textbf{#1}\else #1\fi}

\title{\textbf{Dynamic prediction intervals for survival times}}
\author[1]{Lorenzo Carvisiglia}
\author[1]{Saverio Ranciati}
\author[2]{Mirko Signorelli}
\affil[1]{Department of Statistical Sciences, University of Bologna, Italy}
\affil[2]{Mathematical Institute, Leiden University, The Netherlands}
\date{}

\begin{document}
\maketitle

\begin{abstract}
Most work on survival prediction focuses on estimating survival probabilities rather than predicting individual event times. Recent conformal methods have made it possible to construct prediction intervals for survival times with right-censored outcomes, but existing approaches are restricted to settings with covariates only measured at baseline and do not address dynamic prediction with longitudinal data. We study prediction intervals for survival times in a dynamic prediction framework with longitudinal covariates. Our approach uses Penalized Regression Calibration (PRC) as a working dynamic prediction model, combining linear mixed models for the longitudinal histories with a Cox model for post-landmark survival, and then applies a conformal calibration step to obtain prediction intervals. We compare naive intervals obtained by direct inversion of the survival function estimated by PRC to our dynamic conformal method. A Monte Carlo simulation study evaluates empirical coverage and interval length across sample sizes, censoring levels, landmark times, and non-proportional hazards (NPH) scenarios. We illustrate the proposed methodology by computing dynamic prediction intervals for the time until a dementia diagnosis in the ADNI dataset. The results show that naive inversion is often unreliable, whereas the proposed dynamic conformal method yields more stable predictive performance.
\end{abstract}
\noindent\textbf{Keywords:}
conformal prediction; dynamic prediction; survival analysis; longitudinal data; prediction intervals; penalized regression calibration.

\section{Introduction}

Survival prediction is a central task in biomedical research, where interest often lies in anticipating the future course of a disease or the timing of a clinical event. In many applications, prediction methods estimate the probability that a subject will remain event-free up to a given time, or they provide a full predicted survival curve. These outputs are useful, but in several settings the more direct question is when the event is likely to occur. Prediction intervals (PIs) for survival times address this question by providing a range of plausible values for an individual future event time.

A simple way to obtain such intervals is to fit a survival model and invert the estimated survival function. This naive strategy is intuitive, but it may perform poorly because it treats the fitted model as if it were the true conditional distribution of the event time and does not fully account for the uncertainty induced by model estimation and individual prediction \citep{lawless:2005}. This limitation is especially relevant in survival analysis, where censoring and model misspecification may affect the tail of the fitted survival curve and, therefore, the resulting interval endpoints.

Conformal prediction provides a general framework for constructing predictive sets by combining a working prediction model with a calibration step based on conformity scores \citep{vovk2009line,lei2018distribution}. Its main appeal is that predictive validity does not require the working model to be correctly specified. This makes conformal methods attractive in survival settings, where useful predictions may be obtained from semiparametric or flexible learning methods even when their assumptions are only approximate.

Recent work has extended conformal prediction to censored survival outcomes. \citet{candes:2023} considered right-censored survival data under a Type I censoring design, where censoring occurs at a study end time known for every subject. This setting is useful methodologically, but it differs from the general right-censored settings typically encountered in biomedical applications. Subsequent work has broadened the scope of conformal survival analysis, including methods based on weighted calibration and methods for general right-censored data \citep{gui:2023,sesia:2025,davidov:2025,farina:2025,yi:2025}. Most of these contributions are formulated for settings in which predictions are computed using only covariates measured at baseline. This is restrictive in many biomedical studies, where repeated measurements collected during follow-up may carry substantial information on future survival. Biomarkers, laboratory measurements, cognitive scores, and other clinical variables often evolve over time, and clinically relevant predictions should be updated as new observations become available. This is the setting of dynamic prediction.

Several model-based landmarking methods have recently been proposed for dynamic prediction with numerous longitudinal covariates (see \citet{signorelli:2025benchmark} for a review). These methods make it possible to incorporate information observed up to a landmark time into post-landmark survival prediction, but they output survival probabilities or related risk measures, not PIs for event times. Thus, there remains a need for methods that translate dynamic survival predictions into calibrated PIs for event times, especially in applications involving many longitudinal predictors measured irregularly over time.

In this paper, we address this problem by combining conformal survival prediction with a model-based landmarking approach. We adopt Penalized Regression Calibration (PRC), introduced by \citet{signorelli2021penalized}, as the working dynamic survival model. PRC is attractive here because it was developed for survival prediction with numerous longitudinal predictors and irregular observation times. In the first step, the longitudinal history observed up to the landmark is summarized through subject-specific random effect predictions from linear mixed models (LMMs). In the second step, these summaries are combined with baseline covariates in a Cox model for the survival time.

Although PRC is used as the working model in our implementation, the conformal calibration step is not specific to PRC. The same construction can be applied with any model-based landmarking method that returns an estimated survival function conditional on the information available at the landmark: the key requirement is that the fitted model allows for a subject-specific survival function that can be evaluated and inverted.

Our goal is to use this working dynamic survival model inside a conformal calibration procedure in order to construct PIs for survival times with longitudinal predictors in general right-censored settings. More precisely, we develop a landmark-specific dynamic conformal method inspired by \citet{qin:2025}, but adapted to the case in which the prediction rule depends on both baseline covariates and longitudinal information observed up to the landmark.

The paper makes two main contributions. First, it provides a conformal calibration strategy for dynamic survival PIs when individual predictions are updated using longitudinal covariates observed over follow-up, with an implementation that is suited to settings involving numerous longitudinal predictors. Second, it provides a way to obtain PIs from model-based landmarking methods by calibrating the probability levels used to invert the fitted post-landmark survival function. We evaluate the finite-sample behaviour of the proposed approach across sample sizes, censoring levels, landmark times, and degrees of non-proportional hazards, comparing it with naive intervals obtained by direct inversion of PRC. 
The rest of the manuscript is organized as follows. Section~2 introduces the proposed dynamic conformal prediction method, including the simulation design and evaluation metrics. Section~3 reports the simulation results. Section~4 presents the ADNI real-data application, and Section~5 concludes.

\section{Methods}
\label{methods}

\subsection{Problem setup and data structure}

We consider a longitudinal study with a right-censored survival outcome. For subject $i=1,\ldots,n$, let $T_i$ be the true event time random variable and $C_i$ the censoring time random variable. We observe
$$
\Tstar_i=\min(T_i,C_i),
\qquad
\delta_i=\I\{T_i\le C_i\},
$$
where $\Tstar_i$ is the observed survival time and $\delta_i$ is the event indicator.

At baseline, $p$ time-fixed covariates are observed and collected in the vector $\boldsymbol{X}_i\in\RR^p$. During follow-up, $q$ longitudinal covariates, indexed by $s=1,\ldots,q$, are repeatedly measured at $m_i$ subject-specific visit times
$$
t_{i1}<t_{i2}<\cdots<t_{im_i}<\Tstar_i,
$$
with corresponding observed values
$$
\boldsymbol{Y}_i(t_{ij})=
\bigl(Y_{1ij},\ldots,Y_{qij}\bigr)^\top.
$$

Given a landmark time $\ell>0$, we define $R_i^\ell=\I\{\Tstar_i>\ell\}$ as the variable indicating whether the subject $i$ is still under observation and event-free at time $\ell$, and the landmark risk set 
$$
\Rset(\ell)=\{i:R_i^\ell=1\},
$$
which is the set of subjects who are available for post-landmark prediction.

For a subject $i\in\Rset(\ell)$, let
$$
\boldsymbol{\mathcal{Y}}_{i,\ell}
=
\bigl\{
\boldsymbol{Y}_i(t_{ij}):t_{ij}\le \ell
\bigr\},
$$
denote the observed longitudinal history up to the landmark, and let $\boldsymbol{\mathcal{Y}}_\ell$ denote a generic realized value of such history.

The post-landmark survival function is
$$
S_{\ell}(t\mid \boldsymbol{x},\boldsymbol{\mathcal{Y}}_\ell)
=
\operatorname{P}(T>t\mid T>\ell,\boldsymbol{X}=\boldsymbol{x},\boldsymbol{\mathcal{Y}}_{i,\ell}=\boldsymbol{\mathcal{Y}}_\ell)
\qquad t\ge \ell.
$$
Our goal is to construct a prediction interval (PI) for the event time of a new subject who is observed and event-free at landmark $\ell$, given the baseline covariates and longitudinal measurements observed up to $\ell$. We denote this interval by
$$
 C_{\ell,\alpha}
\bigl(\boldsymbol{X}_{n+1},\boldsymbol{\mathcal{Y}}_{n+1,\ell}\bigr).
$$
We refer to this interval as \textit{valid} in the observed landmark population if, for any level $\alpha\in(0,1)$,
\begin{equation}
    \label{valid}
    \operatorname{P}\left(
T_{n+1}\in
 C_{\ell,\alpha}
\bigl(\boldsymbol{X}_{n+1},\boldsymbol{\mathcal{Y}}_{n+1,\ell}\bigr)
\mid R_{n+1}^{\ell}=1
\right)
\ge 1-\alpha.
\end{equation}

A naive PI can be obtained by directly inverting a fitted survival function. Let
$\widehat S_\ell(t\mid \boldsymbol{x},\boldsymbol{\mathcal{Y}}_\ell)$ be the estimated survival function obtained using a chosen survival model. Define its quantile as 
$$
\widehat q_{1-\alpha}(\boldsymbol{x},\boldsymbol{\mathcal{Y}}_\ell):=\widehat S_\ell^{-1}(\alpha\mid \boldsymbol{x},\boldsymbol{\mathcal{Y}}_\ell)
=
\inf\left\{
t\ge \ell:
\widehat S_\ell(t\mid \boldsymbol{x},\boldsymbol{\mathcal{Y}}_\ell)\le \alpha
\right\}.
$$
Since a survival function is non-increasing, the naive inversion leads to the following two-sided plug-in interval at level $1-\alpha$:

$$
\widehat C_{\ell,\alpha}^{\mathrm{naive}}
\bigl(\boldsymbol{X}_{n+1},\boldsymbol{\mathcal{Y}}_{n+1,\ell}\bigr)
=
\left[
\widehat q_{\alpha/2}\bigl(\boldsymbol{X}_{n+1},\boldsymbol{\mathcal{Y}}_{n+1,\ell}\bigr),
\widehat q_{1-\alpha/2}\bigl(\boldsymbol{X}_{n+1},\boldsymbol{\mathcal{Y}}_{n+1,\ell}\bigr)
\right].
$$
One-sided naive PIs can be obtained in the same way.
Typically, naive PIs perform poorly because they do not adequately account for parameter estimation uncertainty \citep{lawless:2005}. The dynamic conformal method introduced in Section \ref{algsec} keeps the fitted survival model as a working model, but adds a calibration step designed to improve predictive coverage.

\subsection{Working dynamic prediction model}
\label{workingmodel}
We use Penalized Regression Calibration (PRC, \cite{signorelli2021penalized}) as a working joint LMM-Cox model for the conformal dynamic prediction algorithm. Given a landmark time $\ell$ and risk set $\Rset(\ell)$, PRC combines LMMs for the longitudinal covariates with a Cox model for survival. The first component maps the observed longitudinal history $\boldsymbol{\mathcal{Y}}_{i,\ell}$ onto a finite-dimensional subject-specific summary.

First, PRC fits separately for each longitudinal covariate $s=1,\ldots,q$, a LMM of the form
\begin{equation}
\boldsymbol{Y}_{si}(\ell)
=
\boldsymbol{M}_{si}(\ell)\boldsymbol{\beta}_s
+
\boldsymbol{N}_{si}(\ell)\boldsymbol{u}_{si}
+
\boldsymbol{\varepsilon}_{si},
\label{eq:lmm_general_landmark}
\end{equation}
where $\boldsymbol{\beta}_s$ is a vector of fixed effects, $\boldsymbol{u}_{si}$ a vector of subject-specific random effects, $\boldsymbol{M}_{si}(\ell)$ and $\boldsymbol{N}_{si}(\ell)$ are the fixed effect and random effect design matrices, and $\boldsymbol{\varepsilon}_{si}$ is the residual error vector.

A common specification of \eqref{eq:lmm_general_landmark} is the random intercept-random slope model
\begin{equation}
Y_{sij}
=
\beta_{s0}
+
u_{si0}
+
\beta_{s1}t_{ij}
+
u_{si1}t_{ij}
+
\varepsilon_{sij},
\label{eq:lmm_landmark}
\end{equation}
with
$$
\boldsymbol{u}_{si}
=
(u_{si0},u_{si1})^\top
\sim
\mathcal{N}(\boldsymbol{0},\boldsymbol{\Sigma}_s),
\qquad
\varepsilon_{sij}
\sim
\mathcal{N}(0,\sigma_s^2).
$$

After fitting the mixed models, PRC predicts the subject-specific random effects through the Best Linear Unbiased Predictor:
\begin{equation}
\widehat{\boldsymbol{u}}_{si}(\ell)
=
\widehat{\boldsymbol{\Sigma}}_s
\boldsymbol{N}_{si}(\ell)^\top
\widehat{\boldsymbol{V}}_{si}(\ell)^{-1}
\left\{
\boldsymbol{Y}_{si}(\ell)
-
\boldsymbol{M}_{si}(\ell)\widehat{\boldsymbol{\beta}}_s
\right\},
\label{eq:blup_landmark}
\end{equation}
where
$$
\widehat{\boldsymbol{V}}_{si}(\ell)
=
\boldsymbol{N}_{si}(\ell)
\widehat{\boldsymbol{\Sigma}}_s
\boldsymbol{N}_{si}(\ell)^\top
+
\widehat{\sigma}_s^2\boldsymbol{I}.
$$

For subject $i$, the predicted random effects are stacked into the landmark summary
\begin{equation*}
\widehat{\boldsymbol{Z}}_i(\ell)
=
\Bigl(
\widehat{\boldsymbol{u}}_{1i}(\ell)^\top,
\ldots,
\widehat{\boldsymbol{u}}_{qi}(\ell)^\top
\Bigr)^\top.
\end{equation*}

Equivalently, PRC defines a fitted summary operator
$$
\widehat\Psi_\ell(\boldsymbol{\mathcal{Y}}_{i,\ell})
=
\widehat{\boldsymbol{Z}}_i(\ell),
$$
which maps the observed longitudinal history up to the landmark onto the predicted random effects used by the post-landmark survival model.

After having summarized the longitudinal history, PRC then combines baseline covariates and predicted random effects in a Cox model for the event time that is estimated using penalized maximum likelihood. 
The hazard function is
\begin{equation}
h_i(t\mid \boldsymbol{X}_i,\widehat{\boldsymbol{Z}}_i(\ell))
=
h_{0,\ell}(t)
\exp\left\{
\boldsymbol{\gamma}_{x,\ell}^\top\boldsymbol{X}_i
+
\boldsymbol{\gamma}_{z,\ell}^\top\widehat{\boldsymbol{Z}}_i(\ell)
\right\}
\qquad t\ge \ell,
\label{eq:cox_landmark}
\end{equation}

with the corresponding estimated survival function being
\begin{equation}
\widehat S_{\ell}(t\mid \boldsymbol{x},\boldsymbol{z})
=
\exp\left\{
-
\widehat H_{0,\ell}(t)
\exp\bigl(
\widehat{\boldsymbol{\gamma}}_{x,\ell}^\top\boldsymbol{x}
+
\widehat{\boldsymbol{\gamma}}_{z,\ell}^\top\boldsymbol{z}
\bigr)
\right\},
\label{eq:cox_survival_landmark}
\end{equation}
where
$\widehat H_{0,\ell}(t)
=
\int_{\ell}^{t}
\widehat h_{0,\ell}(u)\,\mathrm{d}u$ is the cumulative hazard.

For a generic longitudinal history $\boldsymbol{\mathcal{Y}}_{\ell}$, we write
$$
\widehat S_{\ell}(t\mid \boldsymbol{x},\boldsymbol{\mathcal{Y}}_{\ell})
:=
\widehat S_{\ell}
\left(
t\mid
\boldsymbol{x},
\widehat\Psi_\ell(\boldsymbol{\mathcal{Y}}_{\ell})
\right),
$$
so that the working survival model can be viewed as a function of the original landmark information. This estimated survival function is the working prediction rule used by the conformal procedure.

\subsection{Conformal dynamic prediction algorithm}
\label{algsec}
Algorithm~\ref{alg:dynamic_event_time_predictions} describes the construction of a dynamic PI for the event time of a new subject who is event-free at landmark $\ell$. We present the algorithm using PRC as the working model, but notice that the algorithm can be applied using any model-based landmarking method that returns an invertible subject-specific survival function.

The procedure first summarizes the longitudinal histories available at the landmark and then uses these summaries, together with the baseline covariates, to estimate the working post-landmark survival function. Step 3 accounts for right-censoring in the calibration sample. For a fixed landmark time $\ell$, define the censoring survival function in the observed landmark population as
$$
G_{\ell}(u)
=
\operatorname{P}(C\ge u\mid \Tstar>\ell),
\qquad u\ge \ell.
$$
Assuming that censoring is independent of the covariates, $G_\ell$ is estimated within the landmark risk set using the Kaplan--Meier estimator $\widehat G_\ell$.

The calibration distribution is based on the joint Cumulative Distribution Function (CDF)
$$
F_\ell(t,\boldsymbol{x},\boldsymbol{z})
=
\operatorname{P}
\left(
T\le t,\,
\boldsymbol{X}\le \boldsymbol{x},\,
\boldsymbol{Z}(\ell)\le \boldsymbol{z}
\mid
\Tstar>\ell
\right).
$$
Because event times are right-censored, we estimate this distribution using inverse probability of censoring weights,
$$
w_i^\ell
=
\frac{R_i^\ell\delta_i}{\widehat G_\ell(\Tstar_i)}.
$$
The resulting normalized empirical distribution is
\begin{equation}
\widehat F_{\ell,n}(t,\boldsymbol{x},\boldsymbol{z})
=
\frac{
\sum_{i=1}^{n}
w_i^\ell
\I\{
\Tstar_i\le t,\,
\boldsymbol{X}_i\le\boldsymbol{x},\,
\widehat{\boldsymbol{Z}}_i(\ell)\le \boldsymbol{z}
\}
}{
\sum_{i=1}^{n} w_i^\ell
},
\label{eq:ipcw_landmark_cdf}
\end{equation}
where vector inequalities are interpreted componentwise. Operationally, drawing from \eqref{eq:ipcw_landmark_cdf} means drawing an observed post-landmark failure with probability proportional to $1/\widehat G_\ell(\Tstar_i)$ among subjects in the landmark risk set with $\delta_i=1$. 
In our construction, the conformity score is defined on the survival-probability scale. For a subject with observed post-landmark event time $t$, baseline covariates $\boldsymbol{x}$, and landmark summary $\boldsymbol{z}$, the score is
$$
U=\widehat S_\ell(t\mid \boldsymbol{x},\boldsymbol{z}).
$$
Thus, the score is the fitted post-landmark survival probability assigned to the realized event time. The conformal calibration step uses the empirical distribution of these scores to replace the nominal survival-probability levels used in the naive inversion.

In Step 4, the algorithm performs a subject-level bootstrap. Each bootstrap dataset is obtained by resampling subjects from the landmark risk set, so each selected subject contributes to the observed survival outcome, baseline covariates, and the full longitudinal profile observed up to the landmark. The working model is then refitted on the bootstrap sample. For each bootstrap replicate, one observed post-landmark failure is drawn from the Inverse Probability of Censoring Weighting (IPCW) empirical distribution and its conformity score is computed on the survival-probability scale. In Step 5, the fitted summary operator is applied to the new subject, empirical quantiles of the bootstrap scores are computed, and the final PI is obtained by inverting the fitted post-landmark survival function at the calibrated probability levels.

\begin{algorithm}[tbp]
\SetAlgoSkip{0.35em}
\caption{Conformal dynamic prediction}
\label{alg:dynamic_event_time_predictions}
\DontPrintSemicolon

\Input{landmark time $\ell>0$; landmarked dataset $\Dland=\{(\Tstar_i,\delta_i,\boldsymbol{X}_i,\boldsymbol{\mathcal{Y}}_{i,\ell}):i\in\Rset(\ell)\}$; miscoverage level $\alpha$; number of bootstrap replicates $B$; interval side}
\Output{A prediction interval $\widehat C_{\ell,\alpha}(\boldsymbol{X}_{n+1},\boldsymbol{\mathcal{Y}}_{n+1,\ell})$}

\BlankLine
\AlgStep{1} Estimate the $q$ mixed models, as described in Section \ref{workingmodel}, on $\Dland$ and denote by $\widehat\Psi_\ell$ operator that summarizes the longitudinal covariates into predicted random effects, so that
$$
\widehat\Psi_\ell(\boldsymbol{\mathcal{Y}}_{i,\ell})
=
\widehat{\boldsymbol{Z}}_i(\ell).
$$

\AlgStep{2} Estimate the Cox model described in Section \ref{workingmodel} by penalized maximum likelihood on $\Dland$ and obtain $\widehat S_{\ell}(t\mid \boldsymbol{x},\boldsymbol{z})$.

\AlgStep{3} Estimate the censoring survival function $G_\ell$ by Kaplan-Meier $\widehat G_\ell$ on the landmark risk set and compute weights
$$
w_i^\ell
=
\frac{R_i^\ell\delta_i}{\widehat G_\ell(\Tstar_i)},
\qquad
\pi_i^\ell
=
\frac{w_i^\ell}{\sum_{j=1}^n w_j^\ell}.
$$

\BlankLine
\AlgStep{4} \For{$b=1,\dots,B$}{
Draw a bootstrap dataset $\Dland^{(b)}$ by resampling subjects from $\Dland$ with replacement.

Refit the full working model on $\Dland^{(b)}$ and obtain $\widehat S_{\ell}^{(b)}(t\mid \boldsymbol{x},\boldsymbol{z})$.

Draw an index $i_b$ from $\{1,\ldots,n\}$ with probabilities $\{\pi_i^\ell\}_{i=1}^n$ and compute the predicted random effects for subject $i_b$:
$$
\widehat{\boldsymbol{Z}}_{i_b}^{(b)}(\ell)
=
\widehat\Psi_{\ell}^{(b)}(\boldsymbol{\mathcal{Y}}_{i_b,\ell}).
$$

Compute the score
$$
U_b
=
\widehat S_{\ell}^{(b)}
\bigl(\Tstar_{i_b}\mid
\boldsymbol{X}_{i_b},
\widehat{\boldsymbol{Z}}_{i_b}^{(b)}(\ell)
\bigr).
$$
}

\BlankLine
\AlgStep{5} Compute the predicted random effects for the new subject,
$$
\widehat{\boldsymbol{Z}}_{n+1}(\ell)
=
\widehat\Psi_\ell(\boldsymbol{\mathcal{Y}}_{n+1,\ell}).
$$

\BlankLine
\AlgStep{6} \uIf{$\mathrm{side}=\mathrm{two}$}{
Let $L_{\ell,\alpha}$ and $R_{\ell,\alpha}$ be the empirical $\alpha/2$ and $1-\alpha/2$ quantiles of $\{U_b\}_{b=1}^B$.

Return
$$
\widehat C_{\ell,\alpha}
\bigl(\boldsymbol{X}_{n+1},\boldsymbol{\mathcal{Y}}_{n+1,\ell}\bigr)
=
\left[
\widehat S_{\ell}^{-1}
\bigl(R_{\ell,\alpha}\mid
\boldsymbol{X}_{n+1},\widehat{\boldsymbol{Z}}_{n+1}(\ell)\bigr),
\widehat S_{\ell}^{-1}
\bigl(L_{\ell,\alpha}\mid
\boldsymbol{X}_{n+1},\widehat{\boldsymbol{Z}}_{n+1}(\ell)\bigr)
\right].
$$
}
\uElseIf{$\mathrm{side}=\mathrm{lower}$}{
Let $R_{\ell,\alpha}^{\mathrm{one}}$ be the empirical $1-\alpha$ quantile of $\{U_b\}_{b=1}^B$.

Return
$$
\widehat C_{\ell,\alpha}
\bigl(\boldsymbol{X}_{n+1},\boldsymbol{\mathcal{Y}}_{n+1,\ell}\bigr)
=
\left[
\widehat S_{\ell}^{-1}
\bigl(R_{\ell,\alpha}^{\mathrm{one}}\mid
\boldsymbol{X}_{n+1},\widehat{\boldsymbol{Z}}_{n+1}(\ell)\bigr),
\infty
\right].
$$
}
\Else{
Let $L_{\ell,\alpha}^{\mathrm{one}}$ be the empirical $\alpha$ quantile of $\{U_b\}_{b=1}^B$.

Return
$$
\widehat C_{\ell,\alpha}
\bigl(\boldsymbol{X}_{n+1},\boldsymbol{\mathcal{Y}}_{n+1,\ell}\bigr)
=
\left[
\ell,
\widehat S_{\ell}^{-1}
\bigl(L_{\ell,\alpha}^{\mathrm{one}}\mid
\boldsymbol{X}_{n+1},\widehat{\boldsymbol{Z}}_{n+1}(\ell)\bigr)
\right].
$$
}
\end{algorithm}
\newpage 
We now state the main asymptotic validity result for the proposed landmark conformal interval. The regularity conditions and the proofs are given in Section 3 of the supplementary material.

\begin{theorem}[Asymptotic coverage in the observed landmark population]
\label{thm:landmark_conformal_coverage}
Under regularity conditions stated in Section 3 of the Supplementary Material, the conformal prediction set
$\widehat C_{\ell,\alpha}$ satisfies
$$
\lim_{n\to\infty}\operatorname{P}\left[
T_{n+1}\in
\widehat C_{\ell,\alpha}
\bigl(\boldsymbol X_{n+1},
\boldsymbol{\mathcal Y}_{n+1,\ell}\bigr)
\mid R_{n+1}^{\ell}=1
\right]
=
1-\alpha.
$$
\end{theorem}

The conformal procedure retains the fitted post-landmark survival function but replaces the nominal inversion levels with empirical quantiles of the bootstrap scores. Specifically, the naive interval is obtained by inverting $\widehat S_\ell$ at the nominal probability levels, whereas the conformal interval is obtained by inverting the same fitted survival function at the calibrated levels $L_{\ell,\alpha}$ and $R_{\ell,\alpha}$. These levels reflect the uncertainty arising from estimation of the longitudinal summaries and the survival model, while IPCW sampling accounts for right-censoring among post-landmark event times.

\subsection{Simulation design}

We designed a Monte Carlo simulation study to evaluate the finite-sample behaviour of the proposed dynamic conformal method in a dynamic prediction setting with longitudinal covariates and right-censoring. The data-generating mechanism mimics the setup of a longitudinal study in which subjects enter the study with a baseline age between 40 and 90 years, are repeatedly measured at irregular visit times, and may experience a time-to-event outcome during follow-up. Longitudinal covariates are generated from LMMs with random intercepts and random slopes, so that the information available at a landmark time $\boldsymbol{\mathcal{Y}}_{i,\ell}$ depends on both the number and timing of previous measurements.

The event-time model is constructed so that baseline age and selected features of the longitudinal trajectories are associated with survival. Since the working survival model used by PRC contains a Cox component, event times are deliberately generated under non-proportional hazards (NPH). This allows us to evaluate the behaviour of naive inversion and conformal calibration when the working survival model is only an approximation to the true post-landmark event-time distribution. We consider two NPH scenarios. The weak NPH scenario represents a moderate departure from the Cox working model, whereas the strong NPH scenario creates a more challenging setting with stronger dependence between the longitudinal structure and survival.

Two longitudinal predictor settings are considered. In the first one, there are 3 longitudinal predictors, associated with survival through different components of their subject-specific trajectories: both random intercept and random slope for one marker, random intercept only for another marker, and random slope only for the third marker. In the second setting, there are 20 longitudinal predictors, including both informative and non-informative markers. 

Independent right-censoring is generated under three censoring designs, corresponding to low, medium, and high censoring. These designs target increasing censoring levels in the full dataset before landmarking. After landmarking, the realized censoring rate may differ from the initial target because the analysis is restricted to subjects who are still observed and event-free at the landmark. This reduction of the post-landmark risk set is an important feature of the simulation design, since it directly affects the amount of information available for estimating and calibrating the upper tail of the survival distribution.

Interval performance is evaluated at two landmark times, $\ell=2$ and $\ell=4$. The earlier landmark has a larger post-landmark risk set and more observed post-landmark events, whereas the later landmark provides longer longitudinal histories but fewer subjects and events remaining at risk. Training sample sizes are $N\in\{300,1000,1500\}$, the validation sample size is $10{,}000$, and all performance summaries are based on $1{,}000$ Monte Carlo replications. The full data-generating equations, parameter values, censoring calibration details, and additional landmark descriptive summaries are reported in Sections 1 and 2 of the Supplementary Material.

\subsection{Evaluation metrics}
\label{subsec:evaluation_metrics}

We consider three estimated predictive sets. The lower one-sided interval is
$$
\widehat C_{\ell,\alpha}^{L}(\boldsymbol{x},\boldsymbol{\mathcal{Y}}_{\ell})
=
\bigl[
\widehat L_{\ell}(\boldsymbol{x},\boldsymbol{\mathcal{Y}}_{\ell}),
\infty
\bigr],
$$
the upper one-sided interval is
$$
\widehat C_{\ell,\alpha}^{U}(\boldsymbol{x},\boldsymbol{\mathcal{Y}}_{\ell})
=
\bigl[
\ell,
\widehat U_{\ell}(\boldsymbol{x},\boldsymbol{\mathcal{Y}}_{\ell})
\bigr],
$$
and the two-sided interval is
$$
\widehat C_{\ell,\alpha}^{(2)}(\boldsymbol{x},\boldsymbol{\mathcal{Y}}_{\ell})
=
\bigl[
\widehat L_{\ell}^{(2)}(\boldsymbol{x},\boldsymbol{\mathcal{Y}}_{\ell}),
\widehat U_{\ell}^{(2)}(\boldsymbol{x},\boldsymbol{\mathcal{Y}}_{\ell})
\bigr].
$$
In the simulation study, we estimate the empirical coverage of the PIs as the percentage of subjects whose true event time falls in the PI, conditionally on the fact that the individual is still observed and event-free at the landmark ($R^\ell = 1$). For Monte Carlo replication $r$, let
$$
\mathcal{V}_{\ell}^{(r)}
=
\{i:R_{i}^{\ell,(r)}=1\}
$$
denote the validation subjects who are observed and event-free at landmark $\ell$, and let
$n_{\ell}^{(r)}=|\mathcal{V}_{\ell}^{(r)}|$.

The empirical right coverage, corresponding to the lower one-sided interval
$\widehat C_{\ell,\alpha}^{L}(\boldsymbol{x},\boldsymbol{\mathcal{Y}}_{\ell})$, is obtained by computing
$$
\widehat{\mathrm{Cov}}_{\mathrm{right}}^{(r)}(\ell)
=
\frac{1}{n_{\ell}^{(r)}}
\sum_{i\in\mathcal{V}_{\ell}^{(r)}}
\mathbbm{1}
\left\{
T_i^{(r)}
\ge
\widehat L_{\ell}^{(r)}
\bigl(\boldsymbol{X}_i^{(r)},\boldsymbol{\mathcal{Y}}_{i,\ell}^{(r)}\bigr)
\right\}.
$$
This metric evaluates whether the lower endpoint is not larger than the true event time.

The empirical left coverage, corresponding to the upper one-sided interval
$\widehat C_{\ell,\alpha}^{U}(\boldsymbol{x},\boldsymbol{\mathcal{Y}}_{\ell})$, is obtained by computing
$$
\widehat{\mathrm{Cov}}_{\mathrm{left}}^{(r)}(\ell)
=
\frac{1}{n_{\ell}^{(r)}}
\sum_{i\in\mathcal{V}_{\ell}^{(r)}}
\mathbbm{1}
\left\{
T_i^{(r)}
\le
\widehat U_{\ell}^{(r)}
\bigl(\boldsymbol{X}_i^{(r)},\boldsymbol{\mathcal{Y}}_{i,\ell}^{(r)}\bigr)
\right\}.
$$
This metric evaluates whether the upper endpoint is not smaller than the true event time.

The empirical total coverage, corresponding to the two-sided interval
$\widehat C_{\ell,\alpha}^{(2)}(\boldsymbol{x},\boldsymbol{\mathcal{Y}}_{\ell})$, is obtained by computing
$$
\widehat{\mathrm{Cov}}_{\mathrm{total}}^{(r)}(\ell)
=
\frac{1}{n_{\ell}^{(r)}}
\sum_{i\in\mathcal{V}_{\ell}^{(r)}}
\mathbbm{1}
\left\{
\widehat L_{\ell}^{(2,r)}
\bigl(\boldsymbol{X}_i^{(r)},\boldsymbol{\mathcal{Y}}_{i,\ell}^{(r)}\bigr)
\le
T_i^{(r)}
\le
\widehat U_{\ell}^{(2,r)}
\bigl(\boldsymbol{X}_i^{(r)},\boldsymbol{\mathcal{Y}}_{i,\ell}^{(r)}\bigr)
\right\}.
$$

When the fitted survival function does not yield a finite upper endpoint, the two-sided interval is unbounded from the right. We therefore also consider the truncated upper endpoint
$$
\widehat U_{\ell,\mathrm{tr}}^{(2)}
(\boldsymbol{X},\boldsymbol{\mathcal{Y}}_{\ell})
=
\min
\left\{
\widehat U_{\ell}^{(2)}
(\boldsymbol{X},\boldsymbol{\mathcal{Y}}_{\ell}),
\eta_{\ell}
\right\},
$$
where
$$
\eta_{\ell}
=
\max_{i\in\Rset(\ell):\delta_i=1}\Tstar_i
$$
is the largest observed post-landmark event time in the training sample. The empirical truncated coverage is obtained by computing
$$
\widehat{\mathrm{Cov}}_{\mathrm{trunc}}^{(r)}(\ell)
=
\frac{1}{n_{\ell}^{(r)}}
\sum_{i\in\mathcal{V}_{\ell}^{(r)}}
\mathbbm{1}
\left\{
\widehat L_{\ell}^{(2,r)}
\bigl(\boldsymbol{X}_i^{(r)},\boldsymbol{\mathcal{Y}}_{i,\ell}^{(r)}\bigr)
\le
T_i^{(r)}
\le
\widehat U_{\ell,\mathrm{tr}}^{(2,r)}
\bigl(\boldsymbol{X}_i^{(r)},\boldsymbol{\mathcal{Y}}_{i,\ell}^{(r)}\bigr)
\right\}.
$$

The reported coverage values are the averages of these empirical quantities over the Monte Carlo replications.

Interval width is assessed using the mean length of the truncated two-sided interval,
$$
\mathrm{Length}_{\mathrm{two}}(\ell)
=
\mathbb{E}\left[
\widehat U_{\ell,\mathrm{tr}}^{(2)}(\boldsymbol{X},\boldsymbol{\mathcal{Y}}_{\ell})
-
\widehat L_{\ell}^{(2)}(\boldsymbol{X},\boldsymbol{\mathcal{Y}}_{\ell})
\ \middle|\ R^\ell=1
\right].
$$
We also report the mean length of the upper one-sided interval,
$$
\mathrm{Length}_{\mathrm{upper}}(\ell)
=
\mathbb{E}\left[
\widehat U_{\ell}(\boldsymbol{X},\boldsymbol{\mathcal{Y}}_{\ell})-\ell
\ \middle|\ R^\ell=1
\right],
$$
when the behaviour of the upper endpoint is considered separately.

In the simulation study, all these quantities are estimated empirically on the validation sample after restricting attention to subjects in the observed landmark risk set.

\section{Simulation results}

We report the simulation results by separating the three main scenarios considered in the study. Scenario A corresponds to the weak NPH setting with 3 longitudinal predictors. Scenario B corresponds to the strong NPH setting with 3 longitudinal predictors. Scenario C corresponds to the strong NPH setting with 20 longitudinal predictors, including both informative and non-informative markers.

Across these scenarios, conformal calibration improves the coverage behaviour of the dynamic PIs relative to naive inversion, but the extent of the improvement depends on the difficulty of the setting. In easier cases, calibration brings empirical coverage close to the nominal level. In more challenging cases, especially under stronger censoring, later landmarking, and stronger departure from the Cox working model, calibration improves the endpoint behaviour but does not fully recover nominal coverage.

\subsection{Scenario A: weak NPH with 3 longitudinal predictors}

Table~\ref{tab:weak_3pred_results} reports the results for Scenario A, corresponding to the weak NPH setting with 3 longitudinal predictors. In this scenario, the departure from the Cox working model is moderate, and the longitudinal structure is relatively simple.

\begin{table}[p]
\centering
\scriptsize
\setlength{\tabcolsep}{3pt}
\renewcommand{\arraystretch}{1}
\begin{threeparttable}
\caption{Scenario A, weak NPH with 3 longitudinal predictors: naive and conformal interval results for landmarks $\ell=2$ and $\ell=4$.}
\label{tab:weak_3pred_results}
\begin{tabular}{rlrl|ccccc}
\toprule
$N$ & ESS & $\ell$ & Cens. & Right cov. & Left cov. & Total cov. & Trunc. total cov. & Avg. trunc. length\\
\midrule[1.2pt]
\multicolumn{9}{l}{\textbf{Naive method}}\\
\midrule
300  & 288.2   & 2 & Low    & \textbf{0.906} & 0.874          & \textbf{0.897} & \textbf{0.892} & 11.61\\
1000 & 960.9   & 2 & Low    & \textbf{0.907} & \textbf{0.886} & \textbf{0.902} & \textbf{0.900} & 11.78\\
1500 & 1441.6  & 2 & Low    & \textbf{0.908} & \textbf{0.889} & \textbf{0.903} & \textbf{0.901} & 11.85\\
300  & 212.3   & 4 & Low    & \textbf{0.906} & 0.865          & \textbf{0.892} & \textbf{0.886} & 11.35\\
1000 & 707.1   & 4 & Low    & \textbf{0.908} & 0.879          & \textbf{0.897} & \textbf{0.895} & 11.50\\
1500 & 1060.8  & 4 & Low    & \textbf{0.908} & \textbf{0.883} & \textbf{0.898} & \textbf{0.896} & 11.57\\
\addlinespace[0.35em]
300  & 280.8   & 2 & Medium & \textbf{0.909} & 0.764          & \textbf{0.909} & 0.872          & 9.98\\
1000 & 935.6   & 2 & Medium & \textbf{0.912} & 0.791          & \textbf{0.911} & \textbf{0.884} & 10.32\\
1500 & 1403.2  & 2 & Medium & \textbf{0.912} & 0.793          & \textbf{0.912} & \textbf{0.886} & 10.39\\
300  & 194.1   & 4 & Medium & \textbf{0.903} & 0.733          & \textbf{0.906} & 0.854          & 9.42\\
1000 & 647.1   & 4 & Medium & \textbf{0.906} & 0.772          & \textbf{0.910} & 0.871          & 9.84\\
1500 & 969.9   & 4 & Medium & \textbf{0.906} & 0.776          & \textbf{0.910} & 0.873          & 9.91\\
\addlinespace[0.35em]
300  & 271.5   & 2 & High   & \textbf{0.907} & 0.489          & 0.925          & 0.799          & 7.61\\
1000 & 906.2   & 2 & High   & \textbf{0.911} & 0.515          & 0.930          & 0.820          & 8.00\\
1500 & 1359.1  & 2 & High   & \textbf{0.911} & 0.521          & 0.930          & 0.823          & 8.06\\
300  & 173.1   & 4 & High   & 0.879          & 0.366          & 0.925          & 0.758          & 6.52\\
1000 & 577.7   & 4 & High   & \textbf{0.904} & 0.411          & 0.933          & 0.786          & 6.99\\
1500 & 866.5   & 4 & High   & \textbf{0.905} & 0.422          & 0.934          & 0.791          & 7.09\\
\midrule[1.2pt]
\multicolumn{9}{l}{\textbf{Conformal method}}\\
\midrule
300  & 288.2  & 2 & Low    & \textbf{0.896} & 0.875          & \textbf{0.893} & \textbf{0.888} & 11.91\\
1000 & 960.9  & 2 & Low    & \textbf{0.896} & \textbf{0.882} & \textbf{0.892} & \textbf{0.890} & 11.76\\
1500 & 1441.6 & 2 & Low    & \textbf{0.897} & \textbf{0.885} & \textbf{0.892} & \textbf{0.890} & 11.75\\
300  & 212.3  & 4 & Low    & \textbf{0.901} & 0.868          & \textbf{0.892} & \textbf{0.885} & 11.68\\
1000 & 707.1  & 4 & Low    & \textbf{0.901} & 0.875          & \textbf{0.891} & \textbf{0.888} & 11.45\\
1500 & 1060.8 & 4 & Low    & \textbf{0.902} & 0.878          & \textbf{0.891} & \textbf{0.889} & 11.45\\
\addlinespace[0.35em]
300  & 280.8  & 2 & Medium & \textbf{0.903} & 0.780          & \textbf{0.891} & 0.877          & 9.63\\
1000 & 935.6  & 2 & Medium & \textbf{0.904} & 0.801          & \textbf{0.890} & \textbf{0.886} & 9.79\\
1500 & 1403.2 & 2 & Medium & \textbf{0.904} & 0.803          & \textbf{0.890} & \textbf{0.887} & 9.81\\
300  & 194.1  & 4 & Medium & \textbf{0.903} & 0.765          & \textbf{0.888} & 0.863          & 9.05\\
1000 & 647.1  & 4 & Medium & \textbf{0.904} & 0.794          & \textbf{0.888} & 0.875          & 9.23\\
1500 & 969.9  & 4 & Medium & \textbf{0.903} & 0.797          & \textbf{0.886} & 0.875          & 9.22\\
\addlinespace[0.35em]
300  & 271.5  & 2 & High   & \textbf{0.910} & 0.594          & \textbf{0.883} & 0.797          & 7.13\\
1000 & 906.2  & 2 & High   & \textbf{0.912} & 0.617          & \textbf{0.883} & 0.812          & 7.32\\
1500 & 1359.1 & 2 & High   & \textbf{0.912} & 0.623          & \textbf{0.883} & 0.815          & 7.35\\
300  & 173.1  & 4 & High   & \textbf{0.910} & 0.555          & 0.870          & 0.763          & 6.18\\
1000 & 577.7  & 4 & High   & \textbf{0.903} & 0.579          & \textbf{0.880} & 0.782          & 6.38\\
1500 & 866.6  & 4 & High   & \textbf{0.903} & 0.584          & \textbf{0.880} & 0.790          & 6.42\\
\bottomrule
\end{tabular}
\begin{tablenotes}
\footnotesize
\item ESS denotes the average landmark risk set size. Cens. denotes the initial censoring design: Low, Medium, and High correspond to nominal censoring targets of $20\%$, $35\%$, and $50\%$, respectively. Right cov., Left cov., Total cov., and Trunc. total cov. report Monte Carlo averages of the empirical coverage metrics $\widehat{\mathrm{Cov}}_{\mathrm{right}}$, $\widehat{\mathrm{Cov}}_{\mathrm{left}}$, $\widehat{\mathrm{Cov}}_{\mathrm{total}}$, and $\widehat{\mathrm{Cov}}_{\mathrm{trunc}}$ defined in Section~\ref{subsec:evaluation_metrics}. Avg. trunc. length is the Monte Carlo average of the mean truncated two-sided interval length. Bold coverage values are within $0.02$ of the nominal value $0.90$.
\end{tablenotes}
\end{threeparttable}
\end{table}

For both methods, the empirical right coverage remains close to the nominal level across sample sizes, landmarks, and censoring settings, meaning that the lower endpoint is estimated relatively stably. The reason is that the lower endpoint corresponds to an earlier part of the post-landmark event-time distribution, where the fitted survival curve is informed by more observed failures than in the upper tail.

The empirical left coverage shows a different behaviour. It decreases as censoring increases, particularly at $\ell=4$, because the upper endpoint depends on the upper tail of the event-time distribution. This part of the distribution is more difficult to estimate, since fewer events are observed at later times and right-censoring removes information about late event times. Under low censoring, both methods are reasonably close to the nominal target. Under medium and high censoring, conformal calibration improves the upper-endpoint behaviour relative to naive inversion, although it does not fully restore nominal coverage in the most difficult settings.

The empirical total coverage is often closer to the nominal level than the empirical left coverage. Total coverage combines the behaviour of both endpoints, and the stable lower endpoint can compensate for deficiencies in the upper endpoint. For this reason, total coverage should be interpreted together with truncated total coverage.

Truncated total coverage remains close to the nominal level under low censoring, but decreases under medium and high censoring, especially at the later landmark. The average truncated lengths are comparable between the two methods, and conformal intervals are slightly shorter in several of the more difficult settings.

\subsection{Scenario B: strong NPH with 3 longitudinal predictors}

Table~\ref{tab:strong_3pred_results} reports the results for Scenario B, corresponding to the strong NPH setting with 3 longitudinal predictors. Compared with Scenario A, this setting induces a stronger departure from the Cox working model and is therefore more challenging for predictive calibration.

\begin{table}[p]
\centering
\scriptsize
\setlength{\tabcolsep}{3pt}
\renewcommand{\arraystretch}{1}
\begin{threeparttable}
\caption{Scenario B, strong NPH with 3 longitudinal predictors: naive and conformal interval results for landmarks $\ell=2$ and $\ell=4$.}
\label{tab:strong_3pred_results}
\begin{tabular}{rlrl|ccccc}
\toprule
$N$ & ESS & $\ell$ & Cens. & Right cov. & Left cov. & Total cov. & Trunc. total cov. & Avg. trunc. length\\
\midrule[1.2pt]
\multicolumn{9}{l}{\textbf{Naive method}}\\
\midrule
300  & 281.5   & 2 & Low    & \textbf{0.916} & 0.821          & \textbf{0.909} & \textbf{0.887} & 14.94\\
1000 & 937.7   & 2 & Low    & \textbf{0.918} & 0.837          & \textbf{0.913} & \textbf{0.897} & 15.49\\
1500 & 1406.8  & 2 & Low    & \textbf{0.918} & 0.841          & \textbf{0.914} & \textbf{0.899} & 15.59\\
300  & 197.2   & 4 & Low    & \textbf{0.905} & 0.807          & \textbf{0.899} & 0.870          & 15.35\\
1000 & 657.0   & 4 & Low    & \textbf{0.909} & 0.829          & \textbf{0.905} & \textbf{0.884} & 15.91\\
1500 & 985.7   & 4 & Low    & \textbf{0.910} & 0.833          & \textbf{0.905} & \textbf{0.886} & 16.01\\
\addlinespace[0.35em]
300  & 274.1   & 2 & Medium & \textbf{0.917} & 0.640          & \textbf{0.918} & 0.827          & 11.40\\
1000 & 914.3   & 2 & Medium & \textbf{0.920} & 0.659          & 0.922          & 0.843          & 11.91\\
1500 & 1370.9  & 2 & Medium & \textbf{0.920} & 0.662          & 0.923          & 0.847          & 12.05\\
300  & 180.9   & 4 & Medium & \textbf{0.906} & 0.570          & \textbf{0.912} & 0.791          & 10.96\\
1000 & 603.8   & 4 & Medium & \textbf{0.910} & 0.603          & \textbf{0.916} & 0.811          & 11.59\\
1500 & 907.2   & 4 & Medium & \textbf{0.910} & 0.608          & \textbf{0.917} & 0.814          & 11.74\\
\addlinespace[0.35em]
300  & 265.1   & 2 & High   & \textbf{0.913} & 0.408          & 0.927          & 0.729          & 7.88\\
1000 & 882.5   & 2 & High   & \textbf{0.914} & 0.423          & 0.932          & 0.750          & 8.28\\
1500 & 1324.6  & 2 & High   & \textbf{0.914} & 0.428          & 0.932          & 0.754          & 8.38\\
300  & 161.2   & 4 & High   & \textbf{0.920} & 0.263          & 0.928          & 0.652          & 6.67\\
1000 & 534.6   & 4 & High   & 0.923          & 0.290          & 0.937          & 0.682          & 7.18\\
1500 & 804.1   & 4 & High   & 0.924          & 0.297          & 0.938          & 0.689          & 7.31\\
\midrule[1.2pt]
\multicolumn{9}{l}{\textbf{Conformal method}}\\
\midrule
300  & 281.5  & 2 & Low    & \textbf{0.900} & 0.820          & \textbf{0.896} & \textbf{0.914} & 14.96\\
1000 & 937.7  & 2 & Low    & \textbf{0.900} & 0.835          & \textbf{0.896} & \textbf{0.920} & 15.20\\
1500 & 1406.8 & 2 & Low    & \textbf{0.900} & 0.838          & \textbf{0.895} & 0.921          & 15.22\\
300  & 197.2  & 4 & Low    & \textbf{0.896} & 0.807          & \textbf{0.891} & \textbf{0.902} & 15.33\\
1000 & 657.1  & 4 & Low    & \textbf{0.897} & 0.827          & \textbf{0.892} & \textbf{0.911} & 15.52\\
1500 & 985.8  & 4 & Low    & \textbf{0.897} & 0.830          & \textbf{0.891} & \textbf{0.912} & 15.51\\
\addlinespace[0.35em]
300  & 274.1  & 2 & Medium & \textbf{0.907} & 0.666          & \textbf{0.893} & 0.847          & 10.81\\
1000 & 914.3  & 2 & Medium & \textbf{0.908} & 0.687          & \textbf{0.895} & 0.861          & 11.12\\
1500 & 1370.9 & 2 & Medium & \textbf{0.908} & 0.693          & \textbf{0.895} & 0.863          & 11.21\\
300  & 180.9  & 4 & Medium & \textbf{0.906} & 0.619          & \textbf{0.885} & 0.814          & 10.32\\
1000 & 603.9  & 4 & Medium & \textbf{0.908} & 0.650          & \textbf{0.886} & 0.830          & 10.68\\
1500 & 907.3  & 4 & Medium & \textbf{0.908} & 0.654          & \textbf{0.887} & 0.833          & 10.80\\
\addlinespace[0.35em]
300  & 265.1  & 2 & High   & \textbf{0.909} & 0.490          & \textbf{0.888} & 0.747          & 7.36\\
1000 & 882.5  & 2 & High   & \textbf{0.912} & 0.508          & \textbf{0.890} & 0.762          & 7.60\\
1500 & 1324.6 & 2 & High   & \textbf{0.912} & 0.518          & \textbf{0.887} & 0.764          & 7.64\\
300  & 161.2  & 4 & High   & \textbf{0.904} & 0.396          & \textbf{0.884} & 0.677          & 6.32\\
1000 & 534.6  & 4 & High   & \textbf{0.910} & 0.427          & \textbf{0.886} & 0.698          & 6.64\\
1500 & 804.0  & 4 & High   & \textbf{0.911} & 0.439          & \textbf{0.884} & 0.702          & 6.69\\
\bottomrule
\end{tabular}
\begin{tablenotes}
\footnotesize
\item ESS denotes the average landmark risk set size. Cens. denotes the initial censoring design: Low, Medium, and High correspond to nominal censoring targets of $20\%$, $35\%$, and $50\%$, respectively. Right cov., Left cov., Total cov., and Trunc. total cov. report Monte Carlo averages of the empirical coverage metrics $\widehat{\mathrm{Cov}}_{\mathrm{right}}$, $\widehat{\mathrm{Cov}}_{\mathrm{left}}$, $\widehat{\mathrm{Cov}}_{\mathrm{total}}$, and $\widehat{\mathrm{Cov}}_{\mathrm{trunc}}$ defined in Section~\ref{subsec:evaluation_metrics}. Avg. trunc. length is the Monte Carlo average of the mean truncated two-sided interval length. Bold coverage values are within $0.02$ of the nominal value $0.90$.
\end{tablenotes}
\end{threeparttable}
\end{table}

The empirical right coverage remains close to the nominal level for both methods, but naive inversion tends to be slightly conservative, especially under stronger censoring. Conformal calibration partially corrects this overcoverage, with the clearest correction under low censoring.

The empirical left coverage deteriorates more strongly than in Scenario A. This is expected, because the upper endpoint depends on the upper tail of the time distribution, and this tail is both sparsely observed and more affected by misspecification when the NPH departure is strong. The deterioration is most visible for naive inversion under medium and high censoring and at $\ell=4$.

The advantage of conformal calibration is visible in the upper-endpoint behaviour. Across medium and high censoring settings, conformal calibration increases the empirical left coverage relative to naive inversion. However, the improvement is only partial in the most difficult cases. 

The empirical total coverage is often close to or above the nominal level even when empirical left coverage is poor. As in Scenario A, this reflects the fact that total coverage combines both endpoints, and the lower endpoint is comparatively stable. Truncated total coverage gives a stricter assessment of the finite part of the two-sided interval. Under low censoring, conformal calibration gives truncated total coverage values close to the nominal level. Under medium and high censoring, especially at $\ell=4$, truncated total coverage remains below target despite calibration.

Average truncated lengths are generally comparable between the two methods and are sometimes slightly shorter for the conformal intervals. Therefore, the improvement in coverage behaviour is not driven by a uniform increase in interval length.

\subsection{Scenario C: strong NPH with 20 longitudinal predictors}

Table~\ref{tab:strong_20pred_results} reports the results for Scenario C, corresponding to the strong NPH setting with 20 longitudinal predictors. The longitudinal predictors include both informative and non-informative markers. This setting is statistically more challenging than the 3-predictor scenarios because the working PRC model has to summarize a larger number of longitudinal histories, including markers that do not contribute to survival prediction, before fitting the survival model.

\begin{table}[p]
\centering
\scriptsize
\setlength{\tabcolsep}{3pt}
\renewcommand{\arraystretch}{1}
\begin{threeparttable}
\caption{Scenario C, strong NPH with 20 longitudinal predictors: naive and conformal interval results for landmarks $\ell=2$ and $\ell=4$.}
\label{tab:strong_20pred_results}
\begin{tabular}{rlrl|ccccc}
\toprule
$N$ & ESS & $\ell$ & Cens. & Right cov. & Left cov. & Total cov. & Trunc. total cov. & Avg. trunc. length\\
\midrule[1.2pt]
\multicolumn{9}{l}{\textbf{Naive method}}\\
\midrule
300  & 275.1  & 2 & Low    & \covcell{0.927} & \covcell{0.879} & \covcell{0.891} & \covcell{0.869} & 22.26\\
1000 & 916.4  & 2 & Low    & \covcell{0.911} & \covcell{0.943} & \covcell{0.905} & \covcell{0.883} & 22.95\\
1500 & 1374.7 & 2 & Low    & \covcell{0.911} & \covcell{0.923} & \covcell{0.908} & \covcell{0.887} & 23.29\\
300  & 194.7  & 4 & Low    & \covcell{0.935} & \covcell{0.932} & \covcell{0.906} & \covcell{0.878} & 27.77\\
1000 & 648.6  & 4 & Low    & \covcell{0.915} & \covcell{0.912} & \covcell{0.904} & \covcell{0.888} & 29.39\\
1500 & 972.2  & 4 & Low    & \covcell{0.905} & \covcell{0.892} & \covcell{0.941} & \covcell{0.884} & 28.42\\
\addlinespace[0.35em]
300  & 268.8  & 2 & Medium & \covcell{0.937} & \covcell{0.929} & \covcell{0.916} & \covcell{0.806} & 15.39\\
1000 & 895.1  & 2 & Medium & \covcell{0.951} & \covcell{0.939} & \covcell{0.925} & \covcell{0.825} & 16.08\\
1500 & 1342.7 & 2 & Medium & \covcell{0.941} & \covcell{0.943} & \covcell{0.928} & \covcell{0.828} & 16.22\\
300  & 180.2  & 4 & Medium & \covcell{0.914} & \covcell{0.955} & \covcell{0.940} & \covcell{0.765} & 14.86\\
1000 & 600.5  & 4 & Medium & \covcell{0.905} & \covcell{0.959} & \covcell{0.945} & \covcell{0.786} & 16.24\\
1500 & 900.8  & 4 & Medium & \covcell{0.925} & \covcell{0.956} & \covcell{0.941} & \covcell{0.789} & 16.41\\
\addlinespace[0.35em]
300  & 259.6  & 2 & High   & \covcell{0.953} & \covcell{0.957} & \covcell{0.936} & \covcell{0.684} & 9.41\\
1000 & 864.1  & 2 & High   & \covcell{0.914} & \covcell{0.961} & \covcell{0.944} & \covcell{0.707} & 9.95\\
1500 & 1296.3 & 2 & High   & \covcell{0.928} & \covcell{0.962} & \covcell{0.945} & \covcell{0.712} & 10.09\\
300  & 159.6  & 4 & High   & \covcell{0.954} & \covcell{0.993} & \covcell{0.948} & \covcell{0.590} & 7.69\\
1000 & 531.4  & 4 & High   & \covcell{0.956} & \covcell{0.997} & \covcell{0.950} & \covcell{0.615} & 8.36\\
1500 & 797.7  & 4 & High   & \covcell{0.925} & \covcell{0.996} & \covcell{0.951} & \covcell{0.622} & 8.53\\
\midrule[1.2pt]
\multicolumn{9}{l}{\textbf{Conformal method}}\\
\midrule
300  & 275.1  & 2 & Low    & \covcell{0.905} & \covcell{0.886} & \covcell{0.886} & \covcell{0.885} & 22.33\\
1000 & 916.4  & 2 & Low    & \covcell{0.905} & \covcell{0.888} & \covcell{0.890} & \covcell{0.890} & 22.11\\
1500 & 1374.7 & 2 & Low    & \covcell{0.905} & \covcell{0.889} & \covcell{0.891} & \covcell{0.882} & 22.08\\
300  & 194.7  & 4 & Low    & \covcell{0.907} & \covcell{0.891} & \covcell{0.884} & \covcell{0.884} & 26.96\\
1000 & 648.6  & 4 & Low    & \covcell{0.909} & \covcell{0.892} & \covcell{0.881} & \covcell{0.885} & 26.62\\
1500 & 972.2  & 4 & Low    & \covcell{0.907} & \covcell{0.888} & \covcell{0.877} & \covcell{0.888} & 25.46\\
\addlinespace[0.35em]
300  & 268.8  & 2 & Medium & \covcell{0.913} & \covcell{0.886} & \covcell{0.882} & \covcell{0.853} & 14.45\\
1000 & 895.1  & 2 & Medium & \covcell{0.913} & \covcell{0.872} & \covcell{0.886} & \covcell{0.862} & 14.62\\
1500 & 1342.7 & 2 & Medium & \covcell{0.913} & \covcell{0.865} & \covcell{0.885} & \covcell{0.863} & 14.57\\
300  & 180.2  & 4 & Medium & \covcell{0.916} & \covcell{0.863} & \covcell{0.894} & \covcell{0.766} & 14.67\\
1000 & 600.5  & 4 & Medium & \covcell{0.919} & \covcell{0.857} & \covcell{0.868} & \covcell{0.781} & 15.58\\
1500 & 900.8  & 4 & Medium & \covcell{0.918} & \covcell{0.847} & \covcell{0.858} & \covcell{0.775} & 15.22\\
\addlinespace[0.35em]
300  & 259.6  & 2 & High   & \covcell{0.922} & \covcell{0.854} & \covcell{0.891} & \covcell{0.779} & 9.03\\
1000 & 864.1  & 2 & High   & \covcell{0.914} & \covcell{0.850} & \covcell{0.887} & \covcell{0.789} & 9.16\\
1500 & 1296.3 & 2 & High   & \covcell{0.913} & \covcell{0.850} & \covcell{0.886} & \covcell{0.791} & 9.22\\
300  & 160.0  & 4 & High   & \covcell{0.921} & \covcell{0.798} & \covcell{0.933} & \covcell{0.703} & 7.75\\
1000 & 531.3  & 4 & High   & \covcell{0.915} & \covcell{0.748} & \covcell{0.909} & \covcell{0.726} & 8.37\\
1500 & 797.7  & 4 & High   & \covcell{0.919} & \covcell{0.757} & \covcell{0.901} & \covcell{0.729} & 8.47\\
\bottomrule
\end{tabular}
\begin{tablenotes}
\footnotesize
\item ESS denotes the average landmark risk set size. Cens. denotes the initial censoring design: Low, Medium, and High correspond to nominal censoring targets of $20\%$, $35\%$, and $50\%$, respectively. Right cov., Left cov., Total cov., and Trunc. total cov. report Monte Carlo averages of the empirical coverage metrics $\widehat{\mathrm{Cov}}_{\mathrm{right}}$, $\widehat{\mathrm{Cov}}_{\mathrm{left}}$, $\widehat{\mathrm{Cov}}_{\mathrm{total}}$, and $\widehat{\mathrm{Cov}}_{\mathrm{trunc}}$ defined in Section~\ref{subsec:evaluation_metrics}. Avg. trunc. length is the Monte Carlo average of the mean truncated two-sided interval length. Bold coverage values are within $0.02$ of the nominal value $0.90$.
\end{tablenotes}
\end{threeparttable}
\end{table}

For the conformal method, empirical right coverage remains close to the nominal level across most settings. Naive inversion is often conservative under medium and high censoring, suggesting that the lower endpoint is placed too far toward early event times. Conformal calibration reduces this conservativeness and produces lower-endpoint coverage values that are generally closer to the nominal target.

The empirical left coverage behaves differently from Scenario B. Naive inversion is frequently conservative for the upper endpoint, with values often above the nominal level and becoming especially high under high censoring at $\ell=4$. Conformal calibration reduces this excess coverage and gives more balanced endpoint behaviour, especially under low censoring. Under medium and high censoring, however, upper-endpoint estimation remains difficult.

The empirical total coverage is also more balanced after conformal calibration. Naive inversion tends to be conservative under medium and high censoring, whereas conformal calibration generally brings total coverage closer to the nominal level. 
Truncated total coverage remains the most difficult metric in this scenario. It decreases under medium and high censoring, particularly at $\ell=4$, because truncation removes the effect of extremely late or infinite upper endpoints and focuses on the finite part of the interval. Conformal calibration improves truncated total coverage in several challenging settings, with the clearest gains under high censoring, but the values remain below the nominal level when the risk set is small and censoring is severe.
\par 
To assess the sensitivity of interval length to the nominal coverage level, we repeated the medium-censoring setting of Scenario C with $N=1000$ and landmarks $\ell=2$ and $\ell=4$ for
$$
\alpha\in\{0.05,0.10,0.15,0.20\},
$$
corresponding to nominal coverage levels of $95\%$, $90\%$, $85\%$, and $80\%$. \\Table~\ref{tab:alpha_sensitivity_20cov} reports the Monte Carlo mean of the truncated two-sided interval length over 1,000 replications.

\begin{table}[H]
\centering
\scriptsize
\setlength{\tabcolsep}{4pt}
\renewcommand{\arraystretch}{1}
\begin{threeparttable}
\caption{$\alpha$-sensitivity analysis in Scenario C under medium censoring: mean truncated two-sided interval length.}
\label{tab:alpha_sensitivity_20cov}
\begin{tabular}{ccccc}
\toprule
$\ell$ & $\alpha$ & Nominal coverage & Naive & Conformal \\
\midrule
2 & 0.05 & 0.95 & 17.27 & 16.17 \\
2 & 0.10 & 0.90 & 16.08 & 14.62 \\
2 & 0.15 & 0.85 & 14.95 & 13.29 \\
2 & 0.20 & 0.80 & 13.86 & 12.10 \\
\addlinespace[0.35em]
4 & 0.05 & 0.95 & 16.46 & 16.29 \\
4 & 0.10 & 0.90 & 16.24 & 15.58 \\
4 & 0.15 & 0.85 & 15.97 & 14.44 \\
4 & 0.20 & 0.80 & 15.62 & 13.11 \\
\bottomrule
\end{tabular}
\begin{tablenotes}
\footnotesize
\item The table reports the Monte Carlo mean of the truncated two-sided interval length over 1,000 replications. The nominal coverage is $1-\alpha$. Infinite upper endpoints are truncated at $\eta_\ell$, as defined in Section~\ref{subsec:evaluation_metrics}.
\end{tablenotes}
\end{threeparttable}
\end{table}

Across the considered nominal coverage levels, conformal calibration produces shorter average truncated lengths than naive inversion. As expected, both methods show decreasing interval length as the nominal coverage level decreases.

\section{Application to the ADNI dataset}

\subsection{Data and modelling}

We illustrate the proposed method using data from the Alzheimer's Disease Neuroimaging Initiative (ADNI). ADNI is an ongoing longitudinal study started in 2004 with the aim of identifying and validating biomarkers related to the progression of Alzheimer's disease. Participants undergo repeated clinical, cognitive, imaging, and biomarker assessments over follow-up, with visits scheduled at baseline and at several follow-up times thereafter \citep{weiner2010adni}.

We construct PIs for the time until an individual is diagnosed with dementia. Participants who are not diagnosed with dementia during follow-up are treated as right-censored at their last available visit.

Subjects already diagnosed with dementia at baseline, subjects without follow-up information after baseline, and subjects with missing baseline covariate values are excluded. The resulting dataset contains 1,643 subjects, of which 404 experience dementia during follow-up.

We employ the same covariates considered by \cite{signorelli:2025benchmark}, thus using five baseline covariates (age, sex, baseline diagnostic status, number of years of education, and number of apolipoprotein E $\varepsilon 4$ alleles) and 21 longitudinal covariates (the complete list and description can be found in the same paper).

The processed dataset contains 10,186 visit records. The number of visits per subject ranges from 1 to 22, with an average of approximately 6.2 visits per subject. Follow-up ranges from about 3 months to 15.5 years. We consider landmark times at 2, 3, and 4 years from baseline. 
Predictive performance was evaluated using subject-level five-fold cross-validation.

\subsection{Results}

We applied the proposed PRC-based dynamic conformal method to the ADNI data. We compare it with two reference approaches. The first is naive PRC inversion, obtained by directly inverting the survival function estimated by PRC without conformal calibration; this comparison assesses the effect of the calibration step. The second is conformal baseline Cox, which applies the same conformal calibration procedure to a Cox model fitted using only baseline covariates; this comparison assesses the contribution of the longitudinal history to dynamic prediction.

The dynamic PIs based on PRC were computed using Algorithm~\ref{alg:dynamic_event_time_predictions}, with $B=500$ bootstrap samples. For the baseline Cox model, the same conformal calibration procedure was applied, but the working survival model was fitted only with baseline covariates.

Due to the presence of right-censored subjects for which the true event time is unknown, we estimate the coverage among subjects who received a dementia diagnosis. Table~\ref{tab:adni_cv_results} summarizes the cross-validated behaviour of the three interval constructions in the ADNI application.

\begin{table}[H]
\centering
\scriptsize
\setlength{\tabcolsep}{3pt}
\renewcommand{\arraystretch}{1}
\begin{threeparttable}
\caption{Results of application to ADNI data.}
\label{tab:adni_cv_results}
\begin{tabular}{rlcccc}
\toprule
$\ell$ & Method & Upper cov. & Upper length & Inf. upper & Lower trunc. cov. \tabularnewline
\midrule
2 & Naive PRC & 0.964 & 9.971 & 0.807 & 0.718 \tabularnewline
2 & Conformal PRC & \textbf{0.927} & 9.648 & 0.750 & \textbf{0.868} \tabularnewline
2 & Conformal baseline Cox & 0.945 & \textbf{9.603} & 0.659 & 0.800 \tabularnewline
\addlinespace[0.35em]
3 & Naive PRC & 0.974 & 9.364 & 0.862 & 0.682 \tabularnewline
3 & Conformal PRC & \textbf{0.921} & 8.998 & 0.786 & \textbf{0.841} \tabularnewline
3 & Conformal baseline Cox & 0.947 & \textbf{8.965} & 0.700 & 0.821 \tabularnewline
\addlinespace[0.35em]
4 & Naive PRC & 0.981 & 8.530 & 0.888 & 0.689 \tabularnewline
4 & Conformal PRC & 0.951 & 8.328 & 0.842 & \textbf{0.854} \tabularnewline
4 & Conformal baseline Cox & \textbf{0.942} & \textbf{8.301} & 0.735 & 0.796 \tabularnewline
\bottomrule
\end{tabular}
\begin{tablenotes}
\scriptsize
\item The nominal coverage level is $1-\alpha=0.90$. Upper cov. is computed for the upper one-sided interval $[\ell,\widehat U_\ell]$ among held-out subjects with an observed post-landmark dementia diagnosis. Upper length is the mean length of this interval after truncating infinite upper endpoints at the largest observed event time in the corresponding training landmark risk set. Inf. upper is the proportion of upper one-sided intervals with an infinite upper endpoint. Lower trunc. cov. is computed for the interval $[\widehat L_\ell,\eta_\ell]$, where $\eta_\ell$ is the largest observed event time in the corresponding training landmark risk set. Bold values highlight coverage values closer to the nominal level and, for Upper length, the shortest mean interval within each landmark.
\end{tablenotes}
\end{threeparttable}
\end{table}

The proposed conformal PRC method performs favourably relative to naive PRC inversion. For the upper one-sided intervals, conformal PRC gives coverage values closer to the nominal level than naive PRC at all landmarks, while also reducing the mean upper length and the proportion of infinite upper endpoints. The improvement is even more pronounced for the lower truncated intervals, for which coverage increases substantially after conformal calibration. 

The comparison with conformal baseline Cox gives a more nuanced picture. Conformal PRC yields higher lower truncated coverage at all landmarks, suggesting that the longitudinal history contributes useful predictive information beyond baseline covariates alone. Upper coverage is broadly similar between the two conformal approaches, with conformal PRC closer to the nominal level at two of the three landmarks. However, conformal baseline Cox gives slightly shorter upper intervals and a lower proportion of infinite upper endpoints. 

\section{Conclusions}

We proposed a dynamic conformal method for constructing PIs for survival times in the presence of longitudinal covariates and right-censoring. The method begins with the estimation of a conditional survival curve based on a model-based landmarking dynamic prediction method, and then calibrates the survival-probability levels used for inversion using a conformal method. In this way, it preserves the interpretation of model-based landmarking methods \citep{signorelli:2025benchmark} while adding a resampling-based conformal calibration step inspired by \citet{qin:2025}. Although the implementation in this paper uses PRC as the working model, the calibration strategy is not specific to PRC and can in principle be combined with any landmarking method that returns a subject-specific survival function.

The simulation study shows that naive inversion of the fitted survival function can give misleading interval behaviour, especially for the upper endpoint. This difficulty becomes more pronounced under heavier censoring, later landmark times, and stronger departures from the working Cox model. Across the considered settings, conformal calibration improves the coverage behaviour of the dynamic PIs relative to naive inversion, although this improvement does not necessarily imply that empirical coverage always reaches the nominal level. Coverage close to $1-\alpha$ is easier to obtain for some targets and scenarios than for others: lower-endpoint coverage is generally more stable, whereas upper-endpoint and truncated two-sided coverage remain more challenging under heavier censoring, later landmark times, and stronger departures from the working Cox model. Nevertheless, conformal calibration often improves upper-endpoint coverage and truncated two-sided coverage relative to naive inversion, even when nominal coverage is not fully recovered. Importantly, these gains are not obtained by uniformly increasing interval length. In several settings, conformal PIs have average truncated lengths that are comparable to, and sometimes shorter than, those of the naive PIs. This suggests that the calibration step acts by modifying the survival-probability cutoffs used for inversion, rather than simply widening the intervals.

The 20-predictor simulations confirm that the proposed approach remains useful in a higher-dimensional longitudinal setting including both informative and non-informative markers. In this more challenging case, conformal calibration gives more balanced coverage behaviour than naive inversion and often reduces average truncated length. The $\alpha$-sensitivity analysis further indicates that the qualitative comparison between naive and conformal intervals is not driven by the specific nominal $90\%$ coverage level used in the main simulations.

The ADNI application illustrates the practical use of the method in a real longitudinal survival dataset with irregular repeated measurements and substantial right-censoring. Since true event times are not observed for censored subjects, the real-data analysis cannot provide a formal empirical coverage assessment. Nevertheless, the observed-event diagnostics indicate that the proposed conformal PRC method gives coverage values closer to the nominal level than naive PRC inversion, both for the upper one-sided interval and for the lower truncated interval, while also reducing upper interval length and the proportion of infinite upper endpoints. The comparison between conformal PRC and conformal baseline Cox further suggests that incorporating longitudinal information through PRC improves lower truncated coverage, while yielding broadly similar upper coverage. However, conformal baseline Cox gives slightly shorter upper intervals and fewer infinite upper endpoints, indicating that the advantage of PRC in this application is mainly visible in the lower-endpoint diagnostic rather than in upper interval length. At the same time, the large proportion of infinite upper endpoints highlights the intrinsic difficulty of estimating the upper tail of the post-landmark event-time distribution in this real-data setting.

Several extensions are possible. First, future work could consider calibration procedures that allow for covariate-dependent censoring, for example through more flexible estimators of the censoring distribution. Second, the method could be combined with alternative dynamic survival learners, including alternative landmarking or machine-learning-based approaches, provided that they return invertible survival functions.
Third, future work could extend the framework to interval-censored event times, which are common in longitudinal studies where the event status is assessed only at discrete visits. A recent conformal approach for interval-censored outcomes with time-fixed covariates provides a useful starting point \citep{meixide:2025intervals}. Finally, further work is needed to develop formal real-data diagnostics for predictive coverage under censoring, since observed-event summaries are useful descriptively but do not estimate population-level coverage.

\section*{Data availability}

The ADNI data used in this study are available upon request from the Alzheimer's Disease Neuroimaging Initiative database at
\url{http://adni.loni.usc.edu}.
The R code implementing the proposed method, together with the scripts used to reproduce the simulation study and the ADNI analysis, is available at
\url{https://github.com/lorenzocarvisiglia/dynamicConformalSurv}.

\section*{Acknowledgements}
We acknowledge funding from the Dutch Research Council (NWO) - 
Vidi grant ID: \url{https://doi.org/10.61686/SSXQW87450}.
Data used in the preparation of this article were obtained from the Alzheimer's Disease Neuroimaging Initiative (ADNI) database
\url{http://adni.loni.usc.edu}.
The ADNI investigators contributed to the design and implementation of ADNI and provided access to the data, but they did not participate in the analysis or writing of this manuscript. A complete list of ADNI investigators is available at
\url{https://adni.loni.usc.edu/wp-content/uploads/how_to_apply/ADNI_Acknowledgement_List.pdf}.

\section*{Conflict of interest}
The authors declare no competing interests.

\bibliographystyle{plainnat}
\bibliography{refs}
\end{document}


\maketitle

\section{Detailed simulation data-generating mechanism}
\label{sec:supp_simulation_dgp}

This section reports the full data-generating mechanism used in the simulation study. The main manuscript gives a descriptive overview of the simulation design, while the present section contains the explicit equations, parameter values, and censoring mechanism used to generate the simulated datasets.

\subsection{Baseline age, visit process, and longitudinal trajectories}

For each subject $i$, baseline age is generated from a truncated normal distribution
\[
A_i=\min\{\max(A_i^{\ast},40),90\},
\]
where
\[
A_i^{\ast}\sim \mathcal{N}(60,10^2).
\]

The visit process is generated independently of survival. The number of visits satisfies
\[
m_i\sim\mathrm{Unif}\{4,5,6,7,8,9\}.
\]
One visit is fixed at time $0$, and the remaining $m_i-1$ visit times are sampled independently from $\mathrm{Unif}(0,5)$ and then sorted. If
\[
0=t_{i1}<t_{i2}<\cdots<t_{im_i},
\]
the age at visit $j$ is
\[
\mathrm{age}_{ij}=A_i+t_{ij}.
\]

Let $N_i(\ell)$ denote the number of measurements observed up to landmark $\ell$. Since one measurement is always observed at time $0$, and the remaining $m_i-1$ visit times are independently uniform,
\[
N_i(\ell)
=
1+
\sum_{j=2}^{m_i}\mathbbm{1}\{t_{ij}\le \ell\}.
\]
Conditional on $m_i$,
\[
\mathbb{E}\{N_i(\ell)\mid m_i\}
=
1+(m_i-1)\frac{\ell}{5},
\qquad 0\le \ell\le 5.
\]
Therefore,
\[
\mathbb{E}\{N_i(\ell)\}
=
1+
\{\mathbb{E}(m_i)-1\}\frac{\ell}{5}
=
1+1.1\ell.
\]
Thus, under this visit generation mechanism, the expected number of repeated measurements available up to the landmark depends only on $\ell$.

In the 3-predictor setting, the longitudinal structure is driven by six random effects,
\[
\boldsymbol{b}_i=
\bigl(
b_{i1}^{(0)},
b_{i1}^{(1)},
b_{i2}^{(0)},
b_{i2}^{(1)},
b_{i3}^{(0)},
b_{i3}^{(1)}
\bigr)^\top,
\]
with
\[
\boldsymbol{b}_i\sim \mathcal{N}_6(\boldsymbol{0},\boldsymbol{\Sigma}),
\qquad
\boldsymbol{\Sigma}=\boldsymbol{D}\boldsymbol{R}\boldsymbol{D}.
\]
Here
\[
\boldsymbol{D}=\mathrm{diag}(1.0,0.20,6.0,0.80,3.5,0.55),
\]
and $\boldsymbol{R}$ is block diagonal with three $2\times2$ blocks, each equal to
\[
\begin{pmatrix}
1 & -0.20\\
-0.20 & 1
\end{pmatrix}.
\]
Thus, the random intercept and random slope are negatively correlated within each longitudinal covariate, whereas random effects from different longitudinal covariates are independent.

The three longitudinal covariates are generated according to
\[
Y_{1ij}
=
10-0.30t_{ij}
+
b_{i1}^{(0)}
+
b_{i1}^{(1)}t_{ij}
+
\varepsilon_{1ij},
\qquad
\varepsilon_{1ij}\sim \mathcal{N}(0,0.8^2),
\]
\[
Y_{2ij}
=
100+1.80t_{ij}
+
b_{i2}^{(0)}
+
b_{i2}^{(1)}t_{ij}
+
\varepsilon_{2ij},
\qquad
\varepsilon_{2ij}\sim \mathcal{N}(0,5.0^2),
\]
and
\[
Y_{3ij}
=
50-1.20t_{ij}
+
b_{i3}^{(0)}
+
b_{i3}^{(1)}t_{ij}
+
\varepsilon_{3ij},
\qquad
\varepsilon_{3ij}\sim \mathcal{N}(0,2.5^2).
\]
Residual errors are independent across subjects, visits, and longitudinal covariates.

\subsection{Event-time model and non-proportional hazards scenarios}

The event time follows a shifted lognormal model,
\[
T_i
=
1+\exp\left(\mu_i^{(g)}+\sigma_g Z_i\right),
\qquad
Z_i\sim \mathcal{N}(0,1),
\]
where
\[
g\in\{\mathrm{weak},\mathrm{strong}\}
\]
denotes the non-proportional hazards scenario and
\[
\mu_i^{(g)}=\log 5+\eta_i^{(g)}.
\]
The linear predictor is
\[
\eta_i^{(g)}
=
\beta_{\mathrm{age}}^{(g)}
\frac{A_i-60}{10}
+
\beta_{1,0}^{(g)}
\frac{b_{i1}^{(0)}}{\mathrm{sd}(b_{\cdot 1}^{(0)})}
+
\beta_{1,1}^{(g)}
\frac{b_{i1}^{(1)}}{\mathrm{sd}(b_{\cdot 1}^{(1)})}
+
\beta_{2,0}^{(g)}
\frac{b_{i2}^{(0)}}{\mathrm{sd}(b_{\cdot 2}^{(0)})}
+
\beta_{3,1}^{(g)}
\frac{b_{i3}^{(1)}}{\mathrm{sd}(b_{\cdot 3}^{(1)})}.
\]
Hence the first longitudinal covariate affects survival through both random intercept and random slope, the second only through the random intercept, and the third only through the random slope.

For the weak non-proportional hazards scenario,
\[
\sigma_g=0.50,
\qquad
\beta_{\mathrm{age}}^{(g)}=0.18,
\qquad
\beta_{1,0}^{(g)}=0.20,
\qquad
\beta_{1,1}^{(g)}=0.18,
\qquad
\beta_{2,0}^{(g)}=0.22,
\qquad
\beta_{3,1}^{(g)}=-0.20.
\]
For the strong non-proportional hazards scenario,
\[
\sigma_g=0.40,
\qquad
\beta_{\mathrm{age}}^{(g)}=0.30,
\qquad
\beta_{1,0}^{(g)}=0.40,
\qquad
\beta_{1,1}^{(g)}=0.36,
\qquad
\beta_{2,0}^{(g)}=0.42,
\qquad
\beta_{3,1}^{(g)}=-0.38.
\]

\subsection{Twenty-predictor setting}

We also consider a setting with 20 longitudinal covariates. The data-generating mechanism is the same as in the 3-predictor setting, except that the longitudinal predictors are divided into four groups of equal size. Five predictors are associated with survival through both random intercept and random slope, five through the random intercept only, five through the random slope only, and five are not associated with survival.

The non-zero association coefficients are set according to the corresponding non-proportional hazards scenario. No additional structural changes are made to the landmarking or interval construction procedures. 

\subsection{Censoring mechanism and observed data}

Independent censoring times are generated as
\[
C_i\sim\mathrm{Unif}(1,c_{\max}),
\]
where $c_{\max}$ is calibrated empirically, separately for each non-proportional hazards scenario and target censoring level, in order to achieve approximately $20\%$, $35\%$, or $50\%$ censoring on the original full dataset before landmarking. These targets correspond to the low, medium, and high censoring designs used in the main manuscript.

Longitudinal observations are retained only up to the observed follow-up time, so only visits with $t_{ij}\le \Tstar_i$ are kept. After landmarking, the realized censoring rate in the reduced dataset may differ from the initial target because the landmark analysis conditions on subjects being observed and event-free at the landmark.

\subsection{Simulation dimensions and landmark summaries}

Training sample sizes are
\[
N\in\{300,1000,1500\}.
\]
The landmark descriptive summaries are computed at
\[
\ell\in\{2,2.5,3,4,4.5\}.
\]
The main interval simulation study evaluates performance at $\ell=2$ and $\ell=4$, corresponding to an earlier and a later dynamic prediction setting. The validation sample size is $10{,}000$, and all simulation summaries are based on $1{,}000$ Monte Carlo replications.

For each landmark, the post-landmark descriptive quantities of interest are:
\begin{enumerate}
\item the effective sample size, namely the average size of the landmark risk set;
\item the average number of post-landmark events in the risk set;
\item the post-landmark censoring rate in the reduced landmark dataset;
\item the average number of repeated measurements available up to the landmark.
\end{enumerate}
These descriptive quantities are reported in the following supplementary tables.

\section{Descriptive summaries}
\label{sec:supp_landmark_descriptive}

\setcounter{table}{0}
\renewcommand{\thetable}{A\arabic{table}}

The tables in this section summarize the landmark datasets used in the simulation study. They report the average landmark risk set size, the number of post-landmark events, the realized post-landmark censoring rate, and the mean number of repeated measurements available before each landmark. Across scenarios, later landmarks provide more repeated measurements per subject, but also reduce the number of subjects at risk and the number of post-landmark events. This trade-off is more pronounced under heavier censoring and in the strong NPH scenario.

\begin{table}[H]
\centering
\begin{threeparttable}
\caption{Weak NPH scenario, 3-predictor setting: effective sample size and post-landmark events across landmarks.}
\label{tab:supp_weak_ess_events}
\begin{tabular}{rrccccc}
\toprule
$N$ & Cens. & $\ell=2$ & $\ell=2.5$ & $\ell=3$ & $\ell=4$ & $\ell=4.5$\\
\midrule
\multicolumn{7}{l}{\textit{Panel A: ESS at landmark}}\\
300  & 20\% & 288.2 & 276.0 & 257.9 & 212.3 & 188.2\\
300  & 35\% & 280.8 & 265.1 & 244.1 & 194.1 & 169.4\\
300  & 50\% & 271.5 & 251.8 & 227.4 & 173.1 & 147.4\\
1000 & 20\% & 960.9 & 920.7 & 860.3 & 707.1 & 627.7\\
1000 & 35\% & 935.6 & 883.3 & 813.1 & 647.1 & 564.7\\
1000 & 50\% & 906.2 & 840.4 & 758.9 & 577.7 & 491.2\\
1500 & 20\% & 1441.6 & 1380.9 & 1290.1 & 1060.8 & 941.7\\
1500 & 35\% & 1403.2 & 1324.8 & 1219.8 & 969.9 & 846.6\\
1500 & 50\% & 1359.1 & 1260.5 & 1138.2 & 866.6 & 736.5\\
\addlinespace
\multicolumn{7}{l}{\textit{Panel B: post-landmark events}}\\
300  & 20\% & 237.8 & 230.4 & 216.9 & 179.5 & 159.1\\
300  & 35\% & 192.7 & 185.6 & 172.8 & 137.8 & 119.5\\
300  & 50\% & 147.7 & 140.9 & 129.0 & 97.1  & 81.1\\
1000 & 20\% & 793.5 & 769.0 & 724.1 & 598.4 & 531.3\\
1000 & 35\% & 643.0 & 619.2 & 576.4 & 459.8 & 398.9\\
1000 & 50\% & 493.6 & 470.9 & 430.9 & 324.6 & 270.5\\
1500 & 20\% & 1189.0 & 1152.0 & 1084.5 & 896.7 & 795.8\\
1500 & 35\% & 965.3 & 929.5  & 865.4 & 689.9 & 598.8\\
1500 & 50\% & 739.7 & 705.6  & 645.7 & 486.6 & 405.6\\
\bottomrule
\end{tabular}
\end{threeparttable}
\end{table}

\begin{table}[H]
\centering
\begin{threeparttable}
\caption{Weak NPH scenario, 3-predictor setting: post-landmark censoring rate and mean repeated measurements across landmarks.}
\label{tab:supp_weak_cens_repeats}
\begin{tabular}{rrccccc}
\toprule
$N$ & Cens. & $\ell=2$ & $\ell=2.5$ & $\ell=3$ & $\ell=4$ & $\ell=4.5$\\
\midrule
\multicolumn{7}{l}{\textit{Panel A: post-landmark censoring rate}}\\
300  & 20\% & 0.17 & 0.17 & 0.16 & 0.15 & 0.15\\
300  & 35\% & 0.31 & 0.30 & 0.29 & 0.29 & 0.29\\
300  & 50\% & 0.46 & 0.44 & 0.43 & 0.44 & 0.45\\
1000 & 20\% & 0.17 & 0.16 & 0.16 & 0.15 & 0.15\\
1000 & 35\% & 0.31 & 0.30 & 0.29 & 0.29 & 0.29\\
1000 & 50\% & 0.46 & 0.44 & 0.43 & 0.44 & 0.45\\
1500 & 20\% & 0.18 & 0.17 & 0.16 & 0.15 & 0.16\\
1500 & 35\% & 0.31 & 0.30 & 0.29 & 0.29 & 0.29\\
1500 & 50\% & 0.46 & 0.44 & 0.43 & 0.44 & 0.45\\
\addlinespace
\multicolumn{7}{l}{\textit{Panel B: Mean repeated measurements}}\\
300  & 20\% & 3.20 & 3.75 & 4.30 & 5.40 & 5.95\\
300  & 35\% & 3.20 & 3.75 & 4.30 & 5.40 & 5.95\\
300  & 50\% & 3.20 & 3.75 & 4.30 & 5.40 & 5.95\\
1000 & 20\% & 3.20 & 3.75 & 4.30 & 5.40 & 5.95\\
1000 & 35\% & 3.20 & 3.75 & 4.30 & 5.40 & 5.95\\
1000 & 50\% & 3.20 & 3.75 & 4.30 & 5.40 & 5.95\\
1500 & 20\% & 3.20 & 3.75 & 4.30 & 5.40 & 5.95\\
1500 & 35\% & 3.20 & 3.75 & 4.30 & 5.40 & 5.95\\
1500 & 50\% & 3.20 & 3.75 & 4.30 & 5.40 & 5.95\\
\bottomrule
\end{tabular}
\end{threeparttable}
\end{table}

\begin{table}[H]
\centering
\begin{threeparttable}
\caption{Strong NPH scenario, 3-predictor setting: effective sample size and post-landmark events across landmarks.}
\label{tab:supp_strong_ess_events}
\begin{tabular}{rrccccc}
\toprule
$N$ & Cens. & $\ell=2$ & $\ell=2.5$ & $\ell=3$ & $\ell=4$ & $\ell=4.5$\\
\midrule
\multicolumn{7}{l}{\textit{Panel A: ESS at landmark}}\\
300  & 20\% & 281.5 & 262.2 & 240.0 & 197.2 & 177.6\\
300  & 35\% & 274.1 & 251.8 & 227.7 & 180.9 & 160.3\\
300  & 50\% & 265.1 & 239.1 & 211.7 & 161.2 & 139.3\\
1000 & 20\% & 937.7 & 873.8 & 800.8 & 657.1 & 591.7\\
1000 & 35\% & 914.3 & 839.9 & 758.8 & 603.9 & 535.5\\
1000 & 50\% & 882.5 & 795.1 & 703.6 & 534.6 & 461.7\\
1500 & 20\% & 1406.8 & 1310.8 & 1202.1 & 985.8 & 888.0\\
1500 & 35\% & 1370.9 & 1260.1 & 1138.9 & 907.3 & 804.7\\
1500 & 50\% & 1324.6 & 1194.2 & 1057.2 & 804.0 & 693.8\\
\addlinespace
\multicolumn{7}{l}{\textit{Panel B: post-landmark events}}\\
300  & 20\% & 229.4 & 214.1 & 195.6 & 159.3 & 142.6\\
300  & 35\% & 184.6 & 169.7 & 152.4 & 117.9 & 102.5\\
300  & 50\% & 139.3 & 125.3 & 109.1 & 78.3  & 65.1\\
1000 & 20\% & 763.8 & 712.8 & 652.1 & 530.1 & 474.3\\
1000 & 35\% & 615.9 & 566.0 & 507.9 & 393.7 & 343.0\\
1000 & 50\% & 465.8 & 418.4 & 364.2 & 260.9 & 216.7\\
1500 & 20\% & 1146.6 & 1070.1 & 979.6 & 795.8 & 712.2\\
1500 & 35\% & 921.7 & 847.5 & 760.9 & 590.4 & 514.9\\
1500 & 50\% & 698.5 & 627.8 & 546.5 & 391.8 & 325.1\\
\bottomrule
\end{tabular}
\end{threeparttable}
\end{table}

\begin{table}[H]
\centering
\begin{threeparttable}
\caption{Strong NPH scenario, 3-predictor setting: post-landmark censoring rate and mean repeated measurements across landmarks.}
\label{tab:supp_strong_cens_repeats}
\begin{tabular}{rrccccc}
\toprule
$N$ & Cens. & $\ell=2$ & $\ell=2.5$ & $\ell=3$ & $\ell=4$ & $\ell=4.5$\\
\midrule
\multicolumn{7}{l}{\textit{Panel A: post-landmark censoring rate}}\\
300  & 20\% & 0.18 & 0.18 & 0.19 & 0.19 & 0.20\\
300  & 35\% & 0.33 & 0.33 & 0.33 & 0.35 & 0.36\\
300  & 50\% & 0.47 & 0.48 & 0.48 & 0.51 & 0.53\\
1000 & 20\% & 0.19 & 0.18 & 0.19 & 0.19 & 0.20\\
1000 & 35\% & 0.33 & 0.33 & 0.33 & 0.35 & 0.36\\
1000 & 50\% & 0.47 & 0.47 & 0.48 & 0.51 & 0.53\\
1500 & 20\% & 0.18 & 0.18 & 0.19 & 0.19 & 0.20\\
1500 & 35\% & 0.33 & 0.33 & 0.33 & 0.35 & 0.36\\
1500 & 50\% & 0.47 & 0.47 & 0.48 & 0.51 & 0.53\\
\addlinespace
\multicolumn{7}{l}{\textit{Panel B: Mean repeated measurements}}\\
300  & 20\% & 3.20 & 3.76 & 4.30 & 5.40 & 5.95\\
300  & 35\% & 3.20 & 3.75 & 4.30 & 5.40 & 5.95\\
300  & 50\% & 3.20 & 3.75 & 4.30 & 5.40 & 5.95\\
1000 & 20\% & 3.20 & 3.75 & 4.30 & 5.40 & 5.95\\
1000 & 35\% & 3.20 & 3.75 & 4.30 & 5.40 & 5.95\\
1000 & 50\% & 3.20 & 3.75 & 4.30 & 5.40 & 5.95\\
1500 & 20\% & 3.20 & 3.75 & 4.30 & 5.40 & 5.95\\
1500 & 35\% & 3.20 & 3.75 & 4.30 & 5.40 & 5.95\\
1500 & 50\% & 3.20 & 3.75 & 4.30 & 5.40 & 5.95\\
\bottomrule
\end{tabular}
\end{threeparttable}
\end{table}

\section{Asymptotic validity of the proposed method}
\subsection{Target population and landmark IPCW consistency}
\label{sec:landmark_ipcw}

Fix a landmark time $\ell$ and a finite horizon $\tau>\ell$. For subject $i$, let
\[
\Tstar_i=\min(T_i,C_i),
\qquad
\delta_i=\I\{T_i\le C_i\},
\qquad
R_i^\ell=\I\{\Tstar_i>\ell\}.
\]
Thus, $R_i^\ell=1$ indicates that subject $i$ is observed and event-free at landmark $\ell$.

Let $\boldsymbol{\mathcal Y}_{i,\ell}$ denote the longitudinal history observed up to $\ell$. The oracle landmark summary is
\[
\boldsymbol Z_i^\circ(\ell)
=
\Psi_\ell^\circ(\boldsymbol{\mathcal Y}_{i,\ell})
\in\RR^{d_Z},
\]
where, in our implementation, $d_Z=2q$ when each longitudinal marker is summarized by a random intercept and a random slope. The operator $\Psi_\ell^\circ$ is evaluated at the probability limit of the first-stage mixed-model parameters. The summary is measurable with respect to information available no later than $\ell$, as required in landmark prediction \citep{vanhouwelingen:2007}. Define
\[
\boldsymbol A_i^\circ(\ell)
=
\bigl(\boldsymbol X_i^\top,\boldsymbol Z_i^\circ(\ell)^\top\bigr)^\top .
\]

The observed-landmark target law $P_\ell^\circ$ is the conditional law of
\[
W_\ell^\circ=(T,\boldsymbol X,\boldsymbol Z^\circ(\ell))
\]
given $R^\ell=1$, restricted to $t\in[\ell,\tau]$. Its CDF is
\[
F_\ell^\circ(t,\boldsymbol x,\boldsymbol z)
=
\Prob\{T\le t,\boldsymbol X\le\boldsymbol x,
\boldsymbol Z^\circ(\ell)\le\boldsymbol z\mid R^\ell=1\},
\qquad t\in[\ell,\tau],
\]
where vector inequalities are componentwise. Equivalently, for every bounded measurable function $\varphi$,
\[
P_\ell^\circ\varphi
=
\E\{\varphi(T,\boldsymbol X,\boldsymbol Z^\circ(\ell))\mid R^\ell=1\}.
\]

For $u\ge\ell$, define the censoring survival function in the observed landmark population by
\[
G_\ell(u)=\Prob(C\ge u\mid R^\ell=1).
\]
This is the landmark analogue of standard IPCW representations of survival distributions \citep{sattendatta2001,flemingharrington1991}. In the algorithm, $G_\ell$ is estimated by the Kaplan--Meier estimator of the censoring distribution within the observed risk set. 

The oracle IPCW weight is
\[
w_i^{\ell,\circ}
=
\frac{R_i^\ell\delta_i}{G_\ell(\Tstar_i)}.
\]
The oracle IPCW empirical law is the probability measure
\[
\widehat P_{\ell,n}^{\circ,\mathrm{or}}\varphi
=
\frac{
 \sum_{i=1}^n
 w_i^{\ell,\circ}
 \varphi(\Tstar_i,\boldsymbol X_i,\boldsymbol Z_i^\circ(\ell))
}{
 \sum_{i=1}^n
 w_i^{\ell,\circ}
}.
\]
The corresponding oracle IPCW empirical CDF is
\[
\widehat F_{\ell,n}^{\circ,\mathrm{or}}(t,\boldsymbol x,\boldsymbol z)
=
\frac{
 \sum_{i=1}^n
 w_i^{\ell,\circ}
 \I\{\Tstar_i\le t,\boldsymbol X_i\le\boldsymbol x,
 \boldsymbol Z_i^\circ(\ell)\le\boldsymbol z\}
}{
 \sum_{i=1}^n
 w_i^{\ell,\circ}
}.
\]

\subsection*{Assumptions}

\begin{assumption}[]
\label{ass:sampling_support}
The observations are iid. The oracle summary $\boldsymbol Z^\circ(\ell)$ is measurable with respect to the history observed up to $\ell$. There exists a finite horizon $\tau>\ell$ such that
\[
p_\ell^R=\Prob(R^\ell=1)>0,
\qquad
\Prob(T\le\tau\mid R^\ell=1)=1.
\]

\end{assumption}

\begin{assumption}[]
\label{ass:post_landmark_censoring}
Within the observed landmark population, post-landmark censoring is independent of the event time and landmark covariates. Specifically, for every $u\in[\ell,\tau]$,
\[
\Prob\{C\ge u\mid T,\boldsymbol X,\boldsymbol Z^\circ(\ell),R^\ell=1\}
=
G_\ell(u).
\]
\end{assumption}

\begin{assumption}[Censoring positivity]
\label{ass:censoring_estimator}
There is a constant $g_0>0$ such that
\[
\inf_{u\in[\ell,\tau]}G_\ell(u)\ge g_0.
\]
\end{assumption}

\begin{assumption}[Continuity of the target CDF]
\label{ass:target_continuity}
The CDF $F_\ell^\circ(t,\boldsymbol x,\boldsymbol z)$ is continuous on $[\ell,\tau]\times\RR^p\times\RR^{d_Z}$.
\end{assumption}

\subsection*{Oracle IPCW result}

\begin{theorem}[Oracle observed-landmark IPCW consistency]
\label{thm:oracle_ipcw}
Under Assumptions~\ref{ass:sampling_support}--\ref{ass:target_continuity},
\[
\sup_{t\in[\ell,\tau]}
\sup_{\boldsymbol x\in\RR^p}
\sup_{\boldsymbol z\in\RR^{d_Z}}
\left|
\widehat F_{\ell,n}^{\circ,\mathrm{or}}(t,\boldsymbol x,\boldsymbol z)
-
F_\ell^\circ(t,\boldsymbol x,\boldsymbol z)
\right|
\overset{P}{\to}0.
\]
Consequently, for every bounded continuous function $\varphi$,
\[
\widehat P_{\ell,n}^{\circ,\mathrm{or}}\varphi
\overset{P}{\to}
P_\ell^\circ\varphi .
\]
\end{theorem}

\begin{proof}
Let
\[
\phi_{t,\boldsymbol x,\boldsymbol z}
=
\frac{R^\ell\delta}{G_\ell(\Tstar)}
\I\{\Tstar\le t,\boldsymbol X\le\boldsymbol x,
\boldsymbol Z^\circ(\ell)\le\boldsymbol z\}.
\]
On $\{R^\ell\delta=1\}$, $\Tstar=T$ and $C\ge T$. By iterated expectation and Assumption~\ref{ass:post_landmark_censoring},
\begin{align*}
\E\{\phi_{t,\boldsymbol x,\boldsymbol z}\}
&=
\E\left[
\frac{R^\ell\I\{C\ge T\}}{G_\ell(T)}
\I\{T\le t,\boldsymbol X\le\boldsymbol x,
\boldsymbol Z^\circ(\ell)\le\boldsymbol z\}
\right]
\\
&=
\E\left[
R^\ell
\I\{T\le t,\boldsymbol X\le\boldsymbol x,
\boldsymbol Z^\circ(\ell)\le\boldsymbol z\}
\frac{
\Prob\{C\ge T\mid T,\boldsymbol X,\boldsymbol Z^\circ(\ell),R^\ell=1\}
}{G_\ell(T)}
\right]
\\
&=
\Prob\{R^\ell=1,T\le t,\boldsymbol X\le\boldsymbol x,
\boldsymbol Z^\circ(\ell)\le\boldsymbol z\}
\\
&=
p_\ell^R F_\ell^\circ(t,\boldsymbol x,\boldsymbol z).
\end{align*}
Similarly,
\[
\E\left\{\frac{R^\ell\delta}{G_\ell(\Tstar)}\right\}=p_\ell^R>0.
\]
The class of indicators indexed by $(t,\boldsymbol x,\boldsymbol z)$ is Glivenko--Cantelli. By Assumption~\ref{ass:censoring_estimator}, the multiplier $R^\ell\delta/G_\ell(\Tstar)$ is bounded by $g_0^{-1}$ on the support in Assumption~\ref{ass:sampling_support}. Multiplication by this fixed bounded measurable weight preserves the Glivenko--Cantelli property \citep{vdvwellner1996,vanderVaart1998}. Therefore the numerator converges uniformly to $p_\ell^R F_\ell^\circ(t,\boldsymbol x,\boldsymbol z)$ and the denominator converges to $p_\ell^R$. The denominator is bounded away from zero in probability, and the uniform CDF convergence follows by Slutsky's theorem for ratios. Since the limiting CDF is continuous, the corresponding probability measures converge weakly. The conclusion for bounded continuous test functions follows from the portmanteau theorem.
\end{proof}

\subsection*{Feasible censoring-weight replacement}

Define the censoring-feasible, summary-oracle IPCW empirical law by
\[
\widehat P_{\ell,n}^{\circ,\widehat G}\varphi
=
\frac{
 \sum_{i=1}^n
 \frac{R_i^\ell\delta_i}{\widehat G_\ell(\Tstar_i)}
 \varphi(\Tstar_i,\boldsymbol X_i,\boldsymbol Z_i^\circ(\ell))
}{
 \sum_{i=1}^n
 \frac{R_i^\ell\delta_i}{\widehat G_\ell(\Tstar_i)}
}.
\]
Let $\widehat F_{\ell,n}^{\circ,\widehat G}$ denote its CDF.

\begin{lemma}[Feasible censoring-weight replacement]
\label{lem:censoring_weight_replacement}
Under Assumptions~\ref{ass:sampling_support}--\ref{ass:censoring_estimator},
\[
\sup_{t\in[\ell,\tau]}
\sup_{\boldsymbol x\in\RR^p}
\sup_{\boldsymbol z\in\RR^{d_Z}}
\left|
\widehat F_{\ell,n}^{\circ,\widehat G}(t,\boldsymbol x,\boldsymbol z)
-
\widehat F_{\ell,n}^{\circ,\mathrm{or}}(t,\boldsymbol x,\boldsymbol z)
\right|
\overset{P}{\to}0.
\]
Consequently, for every bounded continuous function $\varphi$,
\[
\widehat P_{\ell,n}^{\circ,\widehat G}\varphi
-
\widehat P_{\ell,n}^{\circ,\mathrm{or}}\varphi
\overset{P}{\to}0.
\]
\end{lemma}

\begin{proof}
Write the CDFs as ratios of empirical averages,
\[
\widehat F_{\ell,n}^{\circ,\widehat G}(t,\boldsymbol x,\boldsymbol z)
=
\frac{P_n\widehat w^\ell I_{t,\boldsymbol x,\boldsymbol z}^\circ}
{P_n\widehat w^\ell},
\qquad
\widehat F_{\ell,n}^{\circ,\mathrm{or}}(t,\boldsymbol x,\boldsymbol z)
=
\frac{P_nw^{\ell,\circ}I_{t,\boldsymbol x,\boldsymbol z}^\circ}
{P_nw^{\ell,\circ}},
\]
where
\[
\widehat w_i^\ell=\frac{R_i^\ell\delta_i}{\widehat G_\ell(\Tstar_i)},
\qquad
w_i^{\ell,\circ}=\frac{R_i^\ell\delta_i}{G_\ell(\Tstar_i)}.
\]
Since $p_\ell^R>0$, the landmark risk-set size $N_\ell=\sum_iR_i^\ell$ diverges in probability. Under Assumption~\ref{ass:post_landmark_censoring} and Assumption~\ref{ass:censoring_estimator}, standard Kaplan--Meier theory applied to the censoring process within the landmark risk set gives
\[
\sup_{u\in[\ell,\tau]}
|\widehat G_\ell(u)-G_\ell(u)|=o_p(1).
\]
Consequently,
\[
\Delta_{G,n}:=
\sup_{u\in[\ell,\tau]}
|\widehat G_\ell(u)^{-1}-G_\ell(u)^{-1}|=o_p(1),
\]
and, with probability tending to one, $\inf_{u\in[\ell,\tau]}\widehat G_\ell(u)\ge g_0/2$. Uniformly in $(t,\boldsymbol x,\boldsymbol z)$,
\[
\left|
P_n\widehat w^\ell I_{t,\boldsymbol x,\boldsymbol z}^\circ
-
P_nw^{\ell,\circ}I_{t,\boldsymbol x,\boldsymbol z}^\circ
\right|
\le
\Delta_{G,n}P_nR^\ell\delta=o_p(1),
\]
because $P_nR^\ell\delta=O_p(1)$. The same bound with $I_{t,\boldsymbol x,\boldsymbol z}^\circ\equiv1$ gives
\[
P_n\widehat w^\ell-P_nw^{\ell,\circ}=o_p(1).
\]
By Theorem~\ref{thm:oracle_ipcw}, $P_nw^{\ell,\circ}\to p_\ell^R>0$ in probability, and therefore $P_n\widehat w^\ell$ is also bounded away from zero in probability. The uniform ratio convergence follows. The statement for bounded continuous test functions follows from weak convergence of the corresponding probability measures.
\end{proof}

\subsection{Generated covariate replacement for PRC summaries and landmark IPCW weights}
\label{sec:generated_covariates}

The first-stage PRC summary is estimated from the data. We therefore need to replace the oracle summary $\boldsymbol Z_i^\circ(\ell)$ by the feasible summary
\[
\widehat{\boldsymbol Z}_i(\ell)
=
\widehat\Psi_\ell(\boldsymbol{\mathcal Y}_{i,\ell}).
\]
The next conditions make explicit the amount of first-stage stability needed for the IPCW empirical law and for prediction of a new subject. They are standard conditions for two-step estimators with generated covariates \citep{pagan1984,murphy1985,mammen2016} and are compatible with the usual consistency properties of mixed-model estimators and BLUPs under regularity conditions \citep{harville1977,kackar1981,kackar1984,robinson1991blup,jiang1998}.

\subsection*{A sufficient condition for PRC oracle-feasible equivalence}

Let $\Theta_\ell$ denote the parameter space for the first-stage mixed models and let $\theta_\ell^\circ\in\Theta_\ell$ be the probability limit of the first-stage estimator. Write
\[
\Psi_\ell(\boldsymbol{\mathcal Y}_{i,\ell};\theta_\ell)
\]
for the PRC summary map evaluated at parameter value $\theta_\ell$, so that
\[
\boldsymbol Z_i^\circ(\ell)
=
\Psi_\ell(\boldsymbol{\mathcal Y}_{i,\ell};\theta_\ell^\circ),
\qquad
\widehat{\boldsymbol Z}_i(\ell)
=
\Psi_\ell(\boldsymbol{\mathcal Y}_{i,\ell};\widehat\theta_\ell).
\]

\begin{assumption}[First-stage consistency]
\label{ass:first_stage_consistency}
The first-stage estimator satisfies
\[
\|\widehat\theta_\ell-\theta_\ell^\circ\|=o_p(1).
\]
The same condition holds for a subject-level bootstrap refit conditionally on the observed sample, namely
\[
\|\widehat\theta_\ell^\ast-\theta_\ell^\circ\|=o_{P^\ast}(1)
\quad\text{in probability}.
\]
\end{assumption}

\begin{assumption}[]
\label{ass:summary_lipschitz}
There exist a neighbourhood $\mathcal N_\ell$ of $\theta_\ell^\circ$ and a measurable envelope $L_\ell(\boldsymbol{\mathcal Y}_{\ell})$ such that, for all $\theta_1,\theta_2\in\mathcal N_\ell$,
\[
\|\Psi_\ell(\boldsymbol{\mathcal Y}_{\ell};\theta_1)
-
\Psi_\ell(\boldsymbol{\mathcal Y}_{\ell};\theta_2)\|_\infty
\le
L_\ell(\boldsymbol{\mathcal Y}_{\ell})\|\theta_1-\theta_2\|.
\]
Moreover,
\[
\E\{R^\ell L_\ell(\boldsymbol{\mathcal Y}_{\ell})\}<\infty.
\]
For a new landmark subject, $L_\ell(\boldsymbol{\mathcal Y}_{n+1,\ell})=O_p(1)$ conditional on $R_{n+1}^\ell=1$.
\end{assumption}

\begin{proposition}[]
\label{prop:sufficient_prc_closeness}
Under Assumptions~\ref{ass:first_stage_consistency} and~\ref{ass:summary_lipschitz}, for every $\eta>0$,
\[
\frac1n\sum_{i=1}^n
R_i^\ell
\I\{\|\widehat{\boldsymbol Z}_i(\ell)-\boldsymbol Z_i^\circ(\ell)\|_\infty>\eta\}
\overset{P}{\to}0,
\]
and for a new subject,
\[
\Prob\{\|\widehat{\boldsymbol Z}_{n+1}(\ell)-\boldsymbol Z_{n+1}^\circ(\ell)\|_\infty>\eta
\mid R_{n+1}^\ell=1\}
\to0.
\]
The corresponding bootstrap statements also hold conditionally in probability with $\widehat{\boldsymbol Z}_i^\ast(\ell)=\widehat\Psi_\ell^\ast(\boldsymbol{\mathcal Y}_{i,\ell})$.
\end{proposition}

\begin{proof}
On the event $\widehat\theta_\ell\in\mathcal N_\ell$,
\[
\|\widehat{\boldsymbol Z}_i(\ell)-\boldsymbol Z_i^\circ(\ell)\|_\infty
\le
L_\ell(\boldsymbol{\mathcal Y}_{i,\ell})
\|\widehat\theta_\ell-\theta_\ell^\circ\|.
\]
For every $M<\infty$,
\begin{align*}
&\frac1n\sum_{i=1}^nR_i^\ell
\I\{L_\ell(\boldsymbol{\mathcal Y}_{i,\ell})
\|\widehat\theta_\ell-\theta_\ell^\circ\|>\eta\}
\\
&\quad\le
\frac1n\sum_{i=1}^nR_i^\ell\I\{L_\ell(\boldsymbol{\mathcal Y}_{i,\ell})>M\}
+
\I\{M\|\widehat\theta_\ell-\theta_\ell^\circ\|>\eta\}.
\end{align*}
The second term is $o_p(1)$ for fixed $M$. The first term converges in probability to $\E[R^\ell\I\{L_\ell>M\}]$, which tends to zero as $M\to\infty$ by integrability. This proves the average statement. The new-subject statement follows from $L_\ell(\boldsymbol{\mathcal Y}_{n+1,\ell})=O_p(1)$ and first-stage consistency. The bootstrap version follows from the same argument with conditional probability $P^\ast$ and the bootstrap consistency in Assumption~\ref{ass:first_stage_consistency}.
\end{proof}

\subsection*{Generated covariate replacement}

Define the fully feasible landmark IPCW empirical law by
\[
\widehat P_{\ell,n}^{\widehat Z,\widehat G}\varphi
=
\frac{
 \sum_{i=1}^n
 \frac{R_i^\ell\delta_i}{\widehat G_\ell(\Tstar_i)}
 \varphi(\Tstar_i,\boldsymbol X_i,\widehat{\boldsymbol Z}_i(\ell))
}{
 \sum_{i=1}^n
 \frac{R_i^\ell\delta_i}{\widehat G_\ell(\Tstar_i)}
}.
\]
Let $\widehat F_{\ell,n}^{\widehat Z,\widehat G}$ denote its CDF.

\begin{proposition}[Generated covariate replacement]
\label{prop:generated_covariate_replacement}
Under Assumptions~\ref{ass:sampling_support}--\ref{ass:target_continuity},
Assumptions~\ref{ass:first_stage_consistency}--\ref{ass:summary_lipschitz},
and Lemma~\ref{lem:censoring_weight_replacement},
\[
\sup_{t\in[\ell,\tau]}
\sup_{\boldsymbol x\in\RR^p}
\sup_{\boldsymbol z\in\RR^{d_Z}}
\left|
\widehat F_{\ell,n}^{\widehat Z,\widehat G}(t,\boldsymbol x,\boldsymbol z)
-
\widehat F_{\ell,n}^{\circ,\widehat G}(t,\boldsymbol x,\boldsymbol z)
\right|
\overset{P}{\to}0.
\]
Consequently,
\[
\sup_{t\in[\ell,\tau]}
\sup_{\boldsymbol x\in\RR^p}
\sup_{\boldsymbol z\in\RR^{d_Z}}
\left|
\widehat F_{\ell,n}^{\widehat Z,\widehat G}(t,\boldsymbol x,\boldsymbol z)
-
F_\ell^\circ(t,\boldsymbol x,\boldsymbol z)
\right|
\overset{P}{\to}0,
\]
and for every bounded continuous function $\varphi$,
\[
\widehat P_{\ell,n}^{\widehat Z,\widehat G}\varphi
\overset{P}{\to}
P_\ell^\circ\varphi.
\]
\end{proposition}

\begin{proof}
Write
\[
\widehat w_i^\ell
=
\frac{R_i^\ell\delta_i}{\widehat G_\ell(\Tstar_i)}.
\]
The two empirical CDFs have the same denominator, so it is enough to show that
\[
\sup_{t,\boldsymbol x,\boldsymbol z}
\left|
P_n\widehat w^\ell
\left(
\widehat I_{t,\boldsymbol x,\boldsymbol z}
-
I_{t,\boldsymbol x,\boldsymbol z}^{\circ}
\right)
\right|
=o_p(1),
\]
where
\[
\widehat I_{t,\boldsymbol x,\boldsymbol z,i}
=
\I\{\Tstar_i\le t,\boldsymbol X_i\le\boldsymbol x,
\widehat{\boldsymbol Z}_i(\ell)\le \boldsymbol z\},
\]
and
\[
I_{t,\boldsymbol x,\boldsymbol z,i}^{\circ}
=
\I\{\Tstar_i\le t,\boldsymbol X_i\le\boldsymbol x,
\boldsymbol Z_i^\circ(\ell)\le \boldsymbol z\}.
\]

For a set \(A\subseteq\RR^{d_Z}\), define the sup-norm distance from
\(\boldsymbol v\in\RR^{d_Z}\) to \(A\) as
\[
\operatorname{dist}_\infty(\boldsymbol v,A)
=
\inf_{\boldsymbol u\in A}
\|\boldsymbol v-\boldsymbol u\|_\infty,
\qquad
\|\boldsymbol v-\boldsymbol u\|_\infty
=
\max_{1\le j\le d_Z}|v_j-u_j|.
\]
For
\[
O_{\boldsymbol z}
=
\{\boldsymbol v\in\RR^{d_Z}:\boldsymbol v\le\boldsymbol z\},
\]
define the \(\eta\)-neighbourhood of its boundary by
\[
\partial_\eta O_{\boldsymbol z}
=
\left\{
\boldsymbol v\in\RR^{d_Z}:
\operatorname{dist}_\infty(\boldsymbol v,\partial O_{\boldsymbol z})\le\eta
\right\}.
\]
If
\[
\|\widehat{\boldsymbol Z}_i(\ell)-\boldsymbol Z_i^\circ(\ell)\|_\infty
\le \eta
\]
and
\[
\boldsymbol Z_i^\circ(\ell)\notin \partial_\eta O_{\boldsymbol z},
\]
then
\[
\I\{\widehat{\boldsymbol Z}_i(\ell)\le\boldsymbol z\}
=
\I\{\boldsymbol Z_i^\circ(\ell)\le\boldsymbol z\}.
\]
Therefore,
\[
\left|
\I\{\widehat{\boldsymbol Z}_i(\ell)\le\boldsymbol z\}
-
\I\{\boldsymbol Z_i^\circ(\ell)\le\boldsymbol z\}
\right|
\le
\I\left\{
\|\widehat{\boldsymbol Z}_i(\ell)-\boldsymbol Z_i^\circ(\ell)\|_\infty>\eta
\right\}
+
\I\left\{
\boldsymbol Z_i^\circ(\ell)\in \partial_\eta O_{\boldsymbol z}
\right\}.
\]
Since the remaining indicators are bounded by one,
\[
\sup_{t,\boldsymbol x,\boldsymbol z}
\left|
P_n\widehat w^\ell
\left(
\widehat I_{t,\boldsymbol x,\boldsymbol z}
-
I_{t,\boldsymbol x,\boldsymbol z}^{\circ}
\right)
\right|
\]
is bounded by
\[
P_n\widehat w^\ell
\I\left\{
\|\widehat{\boldsymbol Z}(\ell)-\boldsymbol Z^\circ(\ell)\|_\infty>\eta
\right\}
+
\sup_{\boldsymbol z}
P_n\widehat w^\ell
\I\left\{
\boldsymbol Z^\circ(\ell)\in \partial_\eta O_{\boldsymbol z}
\right\}.
\]

By the uniform consistency of \(\widehat G_\ell\) and censoring positivity,
with probability tending to one,
\[
\inf_{u\in[\ell,\tau]}\widehat G_\ell(u)\ge g_0/2.
\]
On this event,
\[
0\le \widehat w^\ell\le 2g_0^{-1}.
\]
Thus, by Assumptions~\ref{ass:first_stage_consistency}--\ref{ass:summary_lipschitz},
\[
P_n\widehat w^\ell
\I\left\{
\|\widehat{\boldsymbol Z}(\ell)-\boldsymbol Z^\circ(\ell)\|_\infty>\eta
\right\}
=o_p(1)
\]
for every fixed \(\eta>0\).

For the boundary term,
\[
\sup_{\boldsymbol z}
P_n\widehat w^\ell
\I\left\{
\boldsymbol Z^\circ(\ell)\in \partial_\eta O_{\boldsymbol z}
\right\}
\le
2g_0^{-1}
\sup_{\boldsymbol z}
P_n
\I\left\{
\boldsymbol Z^\circ(\ell)\in \partial_\eta O_{\boldsymbol z}
\right\}
\]
with probability tending to one. The class of boundary-neighbourhood
indicators is a standard Glivenko--Cantelli class. Hence
\[
\sup_{\boldsymbol z}
P_n
\I\left\{
\boldsymbol Z^\circ(\ell)\in \partial_\eta O_{\boldsymbol z}
\right\}
=
\sup_{\boldsymbol z}
\Prob\left\{
\boldsymbol Z^\circ(\ell)\in \partial_\eta O_{\boldsymbol z}
\right\}
+o_p(1).
\]
By the continuity of the distribution of \(\boldsymbol Z^\circ(\ell)\),
\[
\sup_{\boldsymbol z}
\Prob\left\{
\boldsymbol Z^\circ(\ell)\in \partial_\eta O_{\boldsymbol z}
\right\}
\to 0
\qquad
\text{as } \eta\downarrow0.
\]
Therefore,
\[
\sup_{t,\boldsymbol x,\boldsymbol z}
\left|
P_n\widehat w^\ell
\left(
\widehat I_{t,\boldsymbol x,\boldsymbol z}
-
I_{t,\boldsymbol x,\boldsymbol z}^{\circ}
\right)
\right|
=o_p(1).
\]

Since \(P_n\widehat w^\ell\) is bounded away from zero in probability, the
first claim follows. The second claim follows by the triangle inequality,
Lemma~\ref{lem:censoring_weight_replacement}, and
Theorem~\ref{thm:oracle_ipcw}. Finally, since \(F_\ell^\circ\) is continuous,
the uniform convergence of the CDFs implies weak convergence of the associated
probability measures. Hence, for every bounded continuous function \(\varphi\),
\[
\widehat P_{\ell,n}^{\widehat Z,\widehat G}\varphi
\overset{P}{\to}
P_\ell^\circ\varphi.
\]
\end{proof}

\subsection{Bootstrap score distribution and conformal coverage}
\label{sec:bootstrap_coverage}

The previous sections imply that the feasible IPCW law based on $(\Tstar_i,\boldsymbol X_i,\widehat{\boldsymbol Z}_i(\ell))$ consistently estimates the observed-landmark law $P_\ell^\circ$. We now connect this result to the bootstrap score distribution used for conformal calibration.

Let
\[
\widehat S_{\ell}(t\mid\boldsymbol x,\boldsymbol z)
\]
be the fitted post-landmark survival function from the original sample, and let
\[
\widehat S_{\ell}^{\ast}(t\mid\boldsymbol x,\boldsymbol z)
\]
be the corresponding function after a generic subject-level bootstrap refit. The working survival model may be misspecified; the argument only requires convergence to a deterministic working limit.

\subsection*{Additional assumptions}

\begin{assumption}[Working survival convergence]
\label{ass:survival_convergence}
There exists a deterministic function $S_\ell^\circ(t\mid\boldsymbol x,\boldsymbol z)$ such that
\[
\sup_{t\in[\ell,\tau]}
\sup_{\boldsymbol x,\boldsymbol z}
|\widehat S_{\ell}(t\mid\boldsymbol x,\boldsymbol z)-S_\ell^\circ(t\mid\boldsymbol x,\boldsymbol z)|
\overset{P}{\to}0
\]
and
\[
\sup_{t\in[\ell,\tau]}
\sup_{\boldsymbol x,\boldsymbol z}
|\widehat S_{\ell}^{\ast}(t\mid\boldsymbol x,\boldsymbol z)-S_\ell^\circ(t\mid\boldsymbol x,\boldsymbol z)|
\overset{P^\ast}{\to}0
\quad\text{in probability}.
\]
The suprema are taken over the support of $(\boldsymbol X,\boldsymbol Z^\circ(\ell))\mid R^\ell=1$ and over the feasible neighbourhood induced by Assumption~\ref{ass:summary_lipschitz}.
\end{assumption}

\begin{assumption}[Continuity of the limiting score map]
\label{ass:score_map_continuity}
The map
\[
g_\ell(t,\boldsymbol x,\boldsymbol z)=S_\ell^\circ(t\mid\boldsymbol x,\boldsymbol z)
\]
is $P_\ell^\circ$-almost surely continuous. Moreover, if $\|\boldsymbol z_n-\boldsymbol z\|_\infty\to0$, then
\[
|S_\ell^\circ(t\mid\boldsymbol x,\boldsymbol z_n)-S_\ell^\circ(t\mid\boldsymbol x,\boldsymbol z)|\to0
\]
in $P_\ell^\circ$-probability when evaluated at $(t,\boldsymbol x,\boldsymbol z)=(T,\boldsymbol X,\boldsymbol Z^\circ(\ell))$.
\end{assumption}

\begin{assumption}[]
\label{ass:score_no_atoms}
Let
\[
U_\ell^\circ
=
S_\ell^\circ(T\mid\boldsymbol X,\boldsymbol Z^\circ(\ell)),
\qquad
(T,\boldsymbol X,\boldsymbol Z^\circ(\ell))\sim P_\ell^\circ,
\]
and let $H_\ell(u)=\Prob(U_\ell^\circ\le u\mid R^\ell=1)$. Then
\[
\lim_{\eta\downarrow0}
\sup_{u\in[0,1]}
\Prob(|U_\ell^\circ-u|\le\eta\mid R^\ell=1)=0.
\]
For the chosen two-sided miscoverage level $\alpha$, the generalized quantiles
\[
L_{\ell,\alpha}^\circ=\inf\{u:H_\ell(u)\ge\alpha/2\},
\qquad
R_{\ell,\alpha}^\circ=\inf\{u:H_\ell(u)\ge1-\alpha/2\}
\]
satisfy
\[
H_\ell(L_{\ell,\alpha}^\circ)=\alpha/2,
\qquad
H_\ell(R_{\ell,\alpha}^\circ)=1-\alpha/2.
\]
Analogous regularity is assumed for the one-sided quantiles if one-sided intervals are used.
\end{assumption}

\subsection*{Bootstrap score distribution}

Let $\widehat\Psi_\ell^{\ast}$ and $\widehat S_\ell^{\ast}$ be obtained by refitting the full PRC working model on one subject-level bootstrap dataset. Conditional on the original data, draw an index $I$ from $\{1,\ldots,n\}$ with probabilities
\[
\pi_i^\ell
=
\frac{R_i^\ell\delta_i\widehat G_\ell(\Tstar_i)^{-1}}
{\sum_{j=1}^nR_j^\ell\delta_j\widehat G_\ell(\Tstar_j)^{-1}}.
\]
Define
\[
\widehat{\boldsymbol Z}_{I}^{\ast}(\ell)
=
\widehat\Psi_\ell^{\ast}(\boldsymbol{\mathcal Y}_{I,\ell})
\]
and the corresponding ideal bootstrap score
\[
U^{\ast}
=
\widehat S_{\ell}^{\ast}
\{\Tstar_{I}\mid\boldsymbol X_{I},
\widehat{\boldsymbol Z}_{I}^{\ast}(\ell)\}.
\]
Let
\[
\bar H_{\ell,n}(u)=\Prob^\ast(U^\ast\le u)
\]
be the conditional CDF of this generic bootstrap score, given the original data. In implementation, $\bar H_{\ell,n}$ is approximated by drawing $B$ independent bootstrap scores $U_1,\ldots,U_B$ and using
\[
\widehat H_{\ell,B}(u)=\frac{1}{B}\sum_{b=1}^B\I\{U_b\le u\}.
\]
Conditional on the original data, the Dvoretzky--Kiefer--Wolfowitz inequality gives, for every $\varepsilon>0$,
\[
\Prob^\ast\left(
\sup_{u\in[0,1]}
|\widehat H_{\ell,B}(u)-\bar H_{\ell,n}(u)|>\varepsilon
\right)
\le 2\exp(-2B\varepsilon^2),
\]
using the sharp constant of \citet{massart1990}.

\begin{lemma}[Bootstrap IPCW score consistency]
\label{lem:bootstrap_score_consistency}
Assume the conditions of Proposition~\ref{prop:generated_covariate_replacement}, Proposition~\ref{prop:sufficient_prc_closeness}, Assumption~\ref{ass:survival_convergence}, Assumption~\ref{ass:score_map_continuity}, and Assumption~\ref{ass:score_no_atoms}. Then
\[
\sup_{u\in[0,1]}
|\bar H_{\ell,n}(u)-H_\ell(u)|
\overset{P}{\to}0.
\]
\end{lemma}

\begin{proof}
By the bootstrap part of Proposition~\ref{prop:sufficient_prc_closeness}, the bootstrap summary $\widehat\Psi_\ell^\ast(\boldsymbol{\mathcal Y}_{i,\ell})$ is asymptotically equivalent to $\boldsymbol Z_i^\circ(\ell)$ for the original subjects in the landmark risk set, conditionally in probability. Repeating the argument in Proposition~\ref{prop:generated_covariate_replacement}, with conditional probability $P^\ast$, shows that the IPCW law placing mass $\pi_i^\ell$ on
\[
(\Tstar_i,\boldsymbol X_i,\widehat\Psi_\ell^\ast(\boldsymbol{\mathcal Y}_{i,\ell})),
\qquad i=1,\ldots,n,
\]
converges weakly in probability to $P_\ell^\circ$.

By Assumption~\ref{ass:score_map_continuity} and the continuous mapping theorem \citep{vdvwellner1996,vanderVaart1998}, the distribution of $g_\ell(T,\boldsymbol X,\boldsymbol Z)$ under this IPCW law converges weakly to the distribution of $U_\ell^\circ$. Since Assumption~\ref{ass:score_no_atoms} implies continuity of $H_\ell$, Polya's theorem gives uniform convergence of the corresponding CDFs.

It remains to replace the limiting score map $g_\ell$ by the bootstrap fitted score. Let
\[
D_n^\ast=
\sup_{t,\boldsymbol x,\boldsymbol z}
|\widehat S_{\ell}^{\ast}(t\mid\boldsymbol x,\boldsymbol z)-g_\ell(t,\boldsymbol x,\boldsymbol z)|.
\]
By Assumption~\ref{ass:survival_convergence}, $D_n^\ast=o_{P^\ast}(1)$ in probability. For any fixed $\eta>0$, on the event $\{D_n^\ast\le\eta\}$,
\[
\I\{g_\ell(\Tstar_I,\boldsymbol X_I,\widehat{\boldsymbol Z}_I^\ast(\ell))\le u-\eta\}
\le
\I\{U^\ast\le u\}
\le
\I\{g_\ell(\Tstar_I,\boldsymbol X_I,\widehat{\boldsymbol Z}_I^\ast(\ell))\le u+\eta\}.
\]
After integrating with respect to the conditional bootstrap law,
\begin{align*}
\widehat P_{\ell,n}^\ast\{g_\ell\le u-\eta\}
-
\Prob^\ast(D_n^\ast>\eta)
&\le
\bar H_{\ell,n}(u)
\\
&\le
\widehat P_{\ell,n}^\ast\{g_\ell\le u+\eta\}
+
\Prob^\ast(D_n^\ast>\eta),
\end{align*}
where $\widehat P_{\ell,n}^\ast$ denotes the conditional IPCW law described above. Hence
\begin{align*}
\sup_u|\bar H_{\ell,n}(u)-H_\ell(u)|
&\le
\sup_u|\widehat P_{\ell,n}^\ast\{g_\ell\le u\}-H_\ell(u)|
\\
&\quad+
\sup_u\{H_\ell(u+\eta)-H_\ell(u-\eta)\}
+
\Prob^\ast(D_n^\ast>\eta)+o_p(1).
\end{align*}
The first term is $o_p(1)$ by the conditional weak convergence just established and Polya's theorem. The third term is $o_p(1)$ for every fixed $\eta>0$. The second term can be made arbitrarily small as $\eta\downarrow0$ by Assumption~\ref{ass:score_no_atoms}. This proves the result.
\end{proof}

\begin{lemma}[Ideal bootstrap quantile consistency]
\label{lem:quantile_consistency}
Let
\[
\bar L_{\ell,\alpha}=
\inf\{u:\bar H_{\ell,n}(u)\ge\alpha/2\},
\qquad
\bar R_{\ell,\alpha}=
\inf\{u:\bar H_{\ell,n}(u)\ge1-\alpha/2\}.
\]
Under the conditions of Lemma~\ref{lem:bootstrap_score_consistency},
\[
\bar L_{\ell,\alpha}\overset{P}{\to}L_{\ell,\alpha}^\circ,
\qquad
\bar R_{\ell,\alpha}\overset{P}{\to}R_{\ell,\alpha}^\circ.
\]
\end{lemma}

\begin{proof}
Uniform convergence of $\bar H_{\ell,n}$ to $H_\ell$ follows from Lemma~\ref{lem:bootstrap_score_consistency}. The quantile regularity part of Assumption~\ref{ass:score_no_atoms} then gives consistency of the generalized inverse at levels $\alpha/2$ and $1-\alpha/2$ \citep{vanderVaart1998}.
\end{proof}

\begin{lemma}[Monte Carlo cutoff consistency]
\label{lem:monte_carlo_quantiles}
Let
\[
\widehat L_{\ell,\alpha,B}=\inf\{u:\widehat H_{\ell,B}(u)\ge\alpha/2\},
\qquad
\widehat R_{\ell,\alpha,B}=\inf\{u:\widehat H_{\ell,B}(u)\ge1-\alpha/2\}.
\]
If $B=B_n\to\infty$, then, under the conditions of Lemma~\ref{lem:bootstrap_score_consistency},
\[
\widehat L_{\ell,\alpha,B_n}-\bar L_{\ell,\alpha}=o_p(1),
\qquad
\widehat R_{\ell,\alpha,B_n}-\bar R_{\ell,\alpha}=o_p(1).
\]
If $B$ is fixed, these empirical quantiles are a Monte Carlo approximation to the ideal cutoffs and retain a non-vanishing simulation error.
\end{lemma}

\begin{proof}
The conditional DKW inequality gives
\[
\sup_{u\in[0,1]}|\widehat H_{\ell,B_n}(u)-\bar H_{\ell,n}(u)|=o_{P^\ast}(1)
\quad\text{in probability}
\]
whenever $B_n\to\infty$. The claimed quantile convergence follows from the same generalized-inverse argument used in Lemma~\ref{lem:quantile_consistency}. For fixed $B$, the DKW bound does not vanish, so the implemented cutoffs contain an additional Monte Carlo error.
\end{proof}

\subsection*{Final landmark conformal set}

Define the ideal calibrated landmark prediction set by
\[
\bar C_{\ell,\alpha}(\boldsymbol x,\boldsymbol z)
=
\{t\in[\ell,\tau]:
\bar L_{\ell,\alpha}
\le
\widehat S_{\ell}(t\mid\boldsymbol x,\boldsymbol z)
\le
\bar R_{\ell,\alpha}
\}.
\]
For a longitudinal history observed at the landmark, define
\[
\bar C_{\ell,\alpha}(\boldsymbol x,\boldsymbol{\mathcal Y}_{\ell})
:=
\bar C_{\ell,\alpha}
\{\boldsymbol x,\widehat\Psi_\ell(\boldsymbol{\mathcal Y}_{\ell})\}.
\]
If $t\mapsto\widehat S_{\ell}(t\mid\boldsymbol x,\boldsymbol z)$ is strictly decreasing on the relevant range, then
\[
\bar C_{\ell,\alpha}(\boldsymbol x,\boldsymbol z)
=
\left[
\widehat S_{\ell}^{-1}(\bar R_{\ell,\alpha}\mid\boldsymbol x,\boldsymbol z),
\widehat S_{\ell}^{-1}(\bar L_{\ell,\alpha}\mid\boldsymbol x,\boldsymbol z)
\right],
\]
where
\[
\widehat S_{\ell}^{-1}(u\mid\boldsymbol x,\boldsymbol z)
=
\inf\{t\in[\ell,\tau]:\widehat S_{\ell}(t\mid\boldsymbol x,\boldsymbol z)\le u\}.
\]
For step-function survival estimates, such as Cox--Breslow estimates, the theorem is stated for the set $\{t:\bar L_{\ell,\alpha}\le\widehat S_\ell(t\mid\boldsymbol x,\boldsymbol z)\le\bar R_{\ell,\alpha}\}$. The displayed interval is the corresponding generalized-inverse representation under the chosen endpoint convention.

The implemented set $\widehat C_{\ell,\alpha,B}$ replaces $(\bar L_{\ell,\alpha},\bar R_{\ell,\alpha})$ by the empirical Monte Carlo quantiles $(\widehat L_{\ell,\alpha,B},\widehat R_{\ell,\alpha,B})$ computed from $\widehat H_{\ell,B}$.

\begin{theorem}[Asymptotic coverage in the observed landmark population]
\label{thm:landmark_conformal_coverage}
Assume the conditions of Proposition~\ref{prop:generated_covariate_replacement}, Proposition~\ref{prop:sufficient_prc_closeness}, Assumption~\ref{ass:survival_convergence}, Assumption~\ref{ass:score_map_continuity}, and Assumption~\ref{ass:score_no_atoms}. Then the ideal calibrated set satisfies
\[
\Prob\left[
T_{n+1}\in
\bar C_{\ell,\alpha}
\bigl(\boldsymbol X_{n+1},\boldsymbol{\mathcal Y}_{n+1,\ell}\bigr)
\mid R_{n+1}^\ell=1
\right]
\to
1-\alpha.
\]
If, in addition, $B=B_n\to\infty$, then the implemented Monte Carlo set satisfies
\[
\Prob\left[
T_{n+1}\in
\widehat C_{\ell,\alpha,B_n}
\bigl(\boldsymbol X_{n+1},\boldsymbol{\mathcal Y}_{n+1,\ell}\bigr)
\mid R_{n+1}^\ell=1
\right]
\to
1-\alpha.
\]
With fixed $B$, the second statement holds only up to the additional Monte Carlo approximation error of the empirical bootstrap cutoffs.
\end{theorem}

\begin{proof}
Let
\[
U_{n+1}^\circ
=
S_\ell^\circ(T_{n+1}\mid\boldsymbol X_{n+1},
\boldsymbol Z_{n+1}^\circ(\ell)).
\]
By definition, conditional on $R_{n+1}^\ell=1$, $U_{n+1}^\circ$ has CDF $H_\ell$. Hence, by Assumption~\ref{ass:score_no_atoms},
\[
\Prob\{L_{\ell,\alpha}^\circ\le U_{n+1}^\circ\le R_{\ell,\alpha}^\circ
\mid R_{n+1}^\ell=1\}=1-\alpha.
\]
By Lemma~\ref{lem:quantile_consistency},
\[
\bar L_{\ell,\alpha}\overset{P}{\to}L_{\ell,\alpha}^\circ,
\qquad
\bar R_{\ell,\alpha}\overset{P}{\to}R_{\ell,\alpha}^\circ.
\]
Moreover, Assumption~\ref{ass:survival_convergence} gives
\[
\sup_{t,\boldsymbol x,\boldsymbol z}
|\widehat S_{\ell}(t\mid\boldsymbol x,\boldsymbol z)-S_\ell^\circ(t\mid\boldsymbol x,\boldsymbol z)|=o_p(1),
\]
and Proposition~\ref{prop:sufficient_prc_closeness}, together with Assumption~\ref{ass:score_map_continuity}, gives
\[
S_\ell^\circ(T_{n+1}\mid\boldsymbol X_{n+1},\widehat{\boldsymbol Z}_{n+1}(\ell))
-
S_\ell^\circ(T_{n+1}\mid\boldsymbol X_{n+1},\boldsymbol Z_{n+1}^\circ(\ell))
=o_p(1)
\]
conditional on $R_{n+1}^\ell=1$. Therefore,
\[
\widehat S_{\ell}(T_{n+1}\mid\boldsymbol X_{n+1},\widehat{\boldsymbol Z}_{n+1}(\ell))
-
U_{n+1}^\circ=o_p(1)
\]
conditional on $R_{n+1}^\ell=1$.

For every $\eta>0$, Assumption~\ref{ass:score_no_atoms} implies
\[
\Prob\{U_\ell^\circ\in[L_{\ell,\alpha}^\circ-\eta,L_{\ell,\alpha}^\circ+\eta]
\cup[R_{\ell,\alpha}^\circ-\eta,R_{\ell,\alpha}^\circ+\eta]
\mid R^\ell=1\}
\to0
\quad\text{as }\eta\downarrow0.
\]
Together with the convergence of the fitted score and the two ideal bootstrap quantiles, this gives
\begin{align*}
&\E\left[
\I\left\{
\bar L_{\ell,\alpha}
\le
\widehat S_{\ell}(T_{n+1}\mid\boldsymbol X_{n+1},\widehat{\boldsymbol Z}_{n+1}(\ell))
\le
\bar R_{\ell,\alpha}
\right\}
\mid R_{n+1}^\ell=1
\right]
\\
&\qquad\to
\Prob\{L_{\ell,\alpha}^\circ\le U_{n+1}^\circ\le R_{\ell,\alpha}^\circ
\mid R_{n+1}^\ell=1\}
=1-\alpha.
\end{align*}
The event inside the expectation is exactly
\[
\{T_{n+1}\in\bar C_{\ell,\alpha}(\boldsymbol X_{n+1},\boldsymbol{\mathcal Y}_{n+1,\ell})\}
\]
by definition of the conformal set. This proves the ideal-set claim. The implemented-set claim follows by Lemma~\ref{lem:monte_carlo_quantiles}, because replacing $(\bar L_{\ell,\alpha},\bar R_{\ell,\alpha})$ by $(\widehat L_{\ell,\alpha,B_n},\widehat R_{\ell,\alpha,B_n})$ changes the coverage probability by $o(1)$ under the same no-atoms condition. If $B$ is fixed, Lemma~\ref{lem:monte_carlo_quantiles} does not remove the Monte Carlo error.
\end{proof}